\documentclass[
  reprint,                   
  aps,prb,                   
  superscriptaddress,        
  amsmath,amssymb,           
  floatfix                   
]{revtex4-2}

\usepackage{graphicx}
\usepackage{mathtools}
\usepackage{bm}
\usepackage{braket}
\usepackage{microtype}
\usepackage[acronym]{glossaries}
\usepackage[hidelinks]{hyperref}
\usepackage{cleveref}
\usepackage{amsfonts}
\usepackage{xcolor}
\usepackage{physics}
\usepackage{times}
\hypersetup{colorlinks, allcolors=blue}

\usepackage{float}

\newcommand{\ia}{i\alpha}
\newcommand{\jb}{j\beta}

\newcommand{\colgap}{\hspace{1.4ex}}
\newcommand{\mr}[1]{\mathrm{#1}}
\newcommand{\figref}[2]{Fig.~\hyperref[#1]{\ref*{#1}\,(#2)}}

\usepackage{etoolbox}
\pretocmd{\eqref}{Eq.~}{}{} 

\definecolor{Kristoffercol}{HTML}{005F73}
\definecolor{Kristiancol}{HTML}{BB3E03}
\definecolor{Aslecol}{HTML}{6A4C93}

\definecolor{ForestGreen}{RGB}{34,139,34}

\begin{document}

\title{Spin-dependent quasiparticle lifetimes in altermagnets}

\author{Kristoffer Leraand}
\affiliation{Center for Quantum Spintronics, Department of Physics, Norwegian University of Science and Technology, NO-7491 Trondheim, Norway}

\author{Kristian M{\ae}land}
\affiliation{Institute for Theoretical Physics and Astrophysics, University of W{\"u}rzburg, D-97074 W{\"u}rzburg, Germany}
\affiliation{Würzburg-Dresden Cluster of Excellence ct.qmat, D-97074 W{\"u}rzburg, Germany}

\author{Asle Sudb{\o}}
\email{asle.sudbo@ntnu.no}
\affiliation{Center for Quantum Spintronics, Department of Physics, Norwegian University of Science and Technology, NO-7491 Trondheim, Norway}

\begin{abstract}
We investigate many-body effects on the spin-split electron bands in altermagnets by computing the electron self-energy arising from interactions with magnons, phonons, and hybridized magnon-phonon modes. These interactions lead to band broadening, which can obscure the intrinsic spin-splitting in spectroscopic measurements. We consider a $d$-wave Lieb lattice altermagnet as a representative example. Our results reveal that the spin-splitting remains spectroscopically resolvable and provide theoretical estimates of lifetime effects relevant for experimental detection. 
For electron-magnon coupling, we find a distinct difference between spectral function broadening for up and down spins close to the Fermi surface, which is not present in the case of electron-phonon coupling. We relate it to the spin splitting of the magnon modes in altermagnets. The results, including magneto-elastic coupling, are very similar to the pure magnon case. 
This provides insights into quasiparticle dynamics in altermagnets and contributes to the broader understanding of many-body interactions in spin-split systems. By including the temperature dependence of the self-energies, we also quantify how thermal fluctuations influence the broadening of the electronic states. 
\end{abstract}

\maketitle
\section{Introduction}

Altermagnets represent a recently discovered class of magnetic materials characterized by collinear antiferromagnetic order yet hosting spin-split electronic bands \cite{Smejkal-Sinova-Jungwirth, Smejkal2022Dec, Jungwirth2024LiebRev, BBS23,Agterberg24, Durrnagel2024AMLieb, Kaushal2024AMLieb, osumi2023MnTe, LeePRL2024MnTe, krempasky2024MnTe, reimers2023CrSb, Ding2024CrSb, FZhang, Jiang2024AMLiebExp, Wei2024AMLiebExp, GRSMV25, Regmi2025CoNb4Se8, Dale2024CoNb4Se8ARPES}. In contrast to conventional ferromagnets or antiferromagnets, altermagnets enable spin-splitting without net magnetization, making them particularly attractive for spintronics applications where time-reversal symmetry breaking and spin-dependent transport are crucial \cite{Smejkal2020Hall, Sato2024AMHall, Leiviska2024AMHall, Smejkal2022Dec}.

The spin-splitting of altermagnets has attracted great interest into the study of intrinsic unconventional superconductivity \cite{Smejkal2022Dec, mazin2022notes, BBS23, MBS24, LMS25, Hong2025, Chakraborty2024FFLOAM, Bose2024AMSC, Zhang2024FFLOAM, Rasmussen2025AMSC}.
Phonons and magnons are potential mechanisms of unconventional superconductivity in altermagnets \cite{LMS25, BBS23, MBS24}. 
This motivates a systematic study of many-body corrections, in particular the self-energy, to understand how interactions renormalize and broaden the spin-split bands. 

A key question for experimental observation and device application is whether the spin-splitting of the electronic bands in altermagnets remains resolvable in the presence of many-body interactions, particularly those arising from electron coupling to collective excitations such as magnons and phonons. These interactions can introduce significant lifetime broadening of electronic states, potentially masking the intrinsic spin-splitting in spectroscopic measurements. 

In this work, we compute the many-body electronic self-energies of spin-split electron bands in altermagnets arising from interactions with magnons, phonons, and hybridized magnon-phonon excitations. Our goal is to assess the resulting band broadening and its impact on the observability of spin-splitting. We analyze the temperature dependence of the self-energies to determine how thermal effects influence spectral features and broadening. Our results provide insight into the interplay between spin dynamics and electron correlations in altermagnets. Reference \cite{Eto2025MagnonSE} studies the magnon self-energy due to magnon-magnon scattering in altermagnets. Furthermore, Refs.~\cite{ThingstadEliashberg, Thingstad2024May} consider magnon renormalization due to electron-magnon coupling in antiferromagnets. In our case, we focus on the effect on the electrons and do not consider renormalization of the phonons and magnons explicitly. In principle, our parameters could be tuned to reproduce the experimentally obtained, renormalized phonon and magnon spectra in a candidate material.

Our main findings are as follows. We find a spin-dependent lifetime of quasiparticles in altermagnets. If a spin-up and a spin-down band cross the Fermi level in the same momentum direction, they will have different lifetimes below the Fermi level. Alternatively, the band of a single spin has a different lifetime above and below the Fermi level. The sign of the frequency $\omega$ relative to the Fermi level and the spin of the electron decides which magnon mode dominates the interaction. The two magnon modes in altermagnets are spin split \cite{BBS23, MBS24}, or, equivalently \cite{Consoli2025VojtaChiralSpin}, demonstrate chiral splitting \cite{Kravchuk2025BrinkChiralSpin, Liu2024ChiralExp}. Hence, they can yield different electron-magnon coupling strengths. 

This paper is organized as follows. 
We first introduce the lattice, electron, phonon, and magnon models in Sec.~\ref{sec:Models}. 
In Sec.~\ref{sec:Couplings}, we study how the electrons couple to the in-plane (IP) phonons and the magnons, as well as how the magnons couple to out-of-plane (OOP) phonons, leading to new hybrid quasiparticles. 
Considering the lowest-order relevant Feynman diagram, we introduce and calculate the self-energy in Sec.~\ref{sec:Self-energy}.
To experimentally measure the effects of the self-energy, we propose to use angle-resolved photoemission spectroscopy (ARPES) for which the relevant quantity is the spectral function presented in Sec.~\ref{sec:Spectral function}. We conclude our results in Sec.~\ref{sec:Conclusion}. The appendices give further details of our calculations, specifically of the magnon-phonon coupling in App.~\ref{app:Magnon-phonon coupling derivation}, and the diagonalization of a quadratic bosonic Hamiltonian in App.~\ref{app:Diagonalization of quadratic bosonic Hamiltonian}. During a numerical integration, we will need to solve for nodal curves of a two-dimensional function; the algorithm used to find these curves is given in App.~\ref{App:Nodal lines}. It is interesting to determine which magnon species couple to electrons with different spins, depending on the side of the Fermi level we consider. An analysis of this is given in App.~\ref{app:Determining magnon species}. A table of all numerical parameters used is provided in App.~\ref{app:Numerical values}.

\section{Models}
\label{sec:Models}
\begin{figure}[tb]
    \centering
    \includegraphics[width=0.6\linewidth]{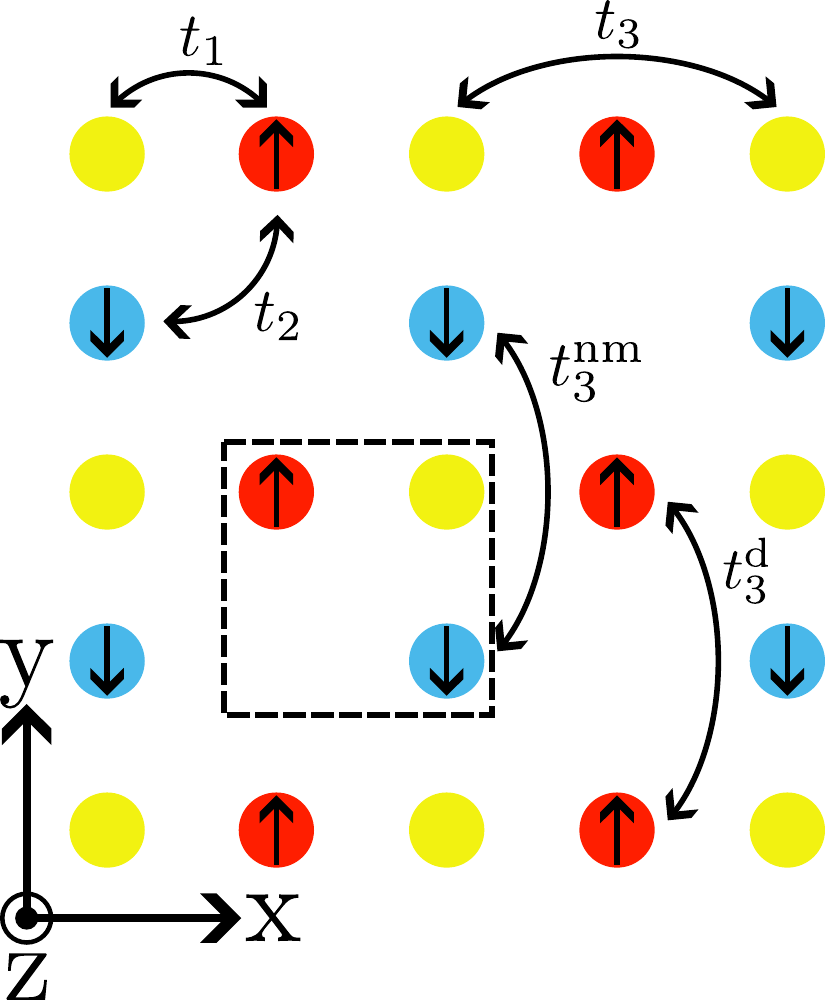}
    \caption{Lieb lattice where the red, blue, and yellow sites have spin-up, spin-down, and no net spin, respectively. The black dashed square marks the unit cell, and the arrows indicate different electron hoppings up to third nearest neighbor hopping.}
    \label{fig:Lieb lattice}
\end{figure}

To model the altermagnet, we consider electrons on the Lieb lattice \cite{Lieb1989Mar}, as shown in Fig.~\ref{fig:Lieb lattice}. Many theory papers use the Lieb lattice to model altermagnets \cite{BBS23, MBS24, LMS25, Antonenko2025Lieb, Durrnagel2024AMLieb, Kaushal2024AMLieb, Petermann2025Lieb, Xu2025Lieb, Jungwirth2024LiebRev, Chang2025Lieb}. Realistic compounds with this lattice structure include $\rm{Rb_{1-\delta} V_2 Te_2 O}$, $\rm{K V_2 Se_2 O}$, and ${\mathrm{La}}_{2}{\mathrm{O}}_{3}{\mathrm{Mn}}_{2}{\mathrm{Se}}_{2}$ \cite{Jiang2024AMLiebExp,FZhang, Wei2024AMLiebExp, GRSMV25, Chang2025Lieb}. We base some parameter values upon ${\mathrm{La}}_{2}{\mathrm{O}}_{3}{\mathrm{Mn}}_{2}{\mathrm{Se}}_{2}$. See App.~\ref{app:Numerical values} for more details.

Ref.~\cite{BBS23} introduced a simple model of a metallic altermagnetic system of itinerant electrons hopping on a Lieb lattice consisting of a unit cell of three atoms. On two of the atomic sites of the unit cell, the model has separate localized spin degrees of freedom ordered in an antiferromagnetic manner. 
Lattice symmetries play a central role in our analysis, particularly for the phonon properties discussed in Sec.~\ref{sec:phonons}, because they determine which vibrational modes of the lattice ions are allowed. For the Lieb lattice illustrated in Fig.~\ref{fig:Lieb lattice}, the symmetry operations are given by
\begin{equation}
\label{eq:Lattice symmetries}
    \begin{aligned}
    \begin{array}{cccc}
        \left(E|0\right), &\left(C_{2z}|0\right), &\left(\sigma_x|0\right), &\left(\sigma_y|0\right),\\[0.5em]
        \left(C_{4z}^+|\mathcal{T}\right), &\left(C_{4z}^-|\mathcal{T}\right), &\left(\sigma_{xy}|\mathcal{T}\right), &\left(\sigma_{\Bar{xy}}|\mathcal{T}\right),
        \end{array}
    \end{aligned}
\end{equation}
where \textit{E} is the identity operation, $C_{\mathrm{2z}}$ is a twofold rotation about the \textit{z}-axis, and $C_{\mathrm{4z}}^\pm$ are fourfold clockwise and counterclockwise rotations, respectively. The operators $\sigma_x$, $\sigma_y$, $\sigma_{xy}$, and $\sigma_{\Bar{xy}}$ represent mirror reflections about the \textit{x}-axis, \textit{y}-axis, and the diagonals $y=x$ and the $y=-x$, respectively. The operator $\mathcal{T}$ denotes time reversal, which flips the on-site spins. The notation $\left(A|0\right)$ indicates a pure symmetry of \textit{A}, while $\left(B|\mathcal{T}\right)$ represents a combined symmetry involving both a spatial operation \textit{B} and time reversal.

In terms of bosons, we allow phonon and magnon collective-mode excitations, as well as a magneto-elastic coupling between magnons and OOP phonons. 

\begin{figure}[tb]
    \centering
    \includegraphics[width=\linewidth]{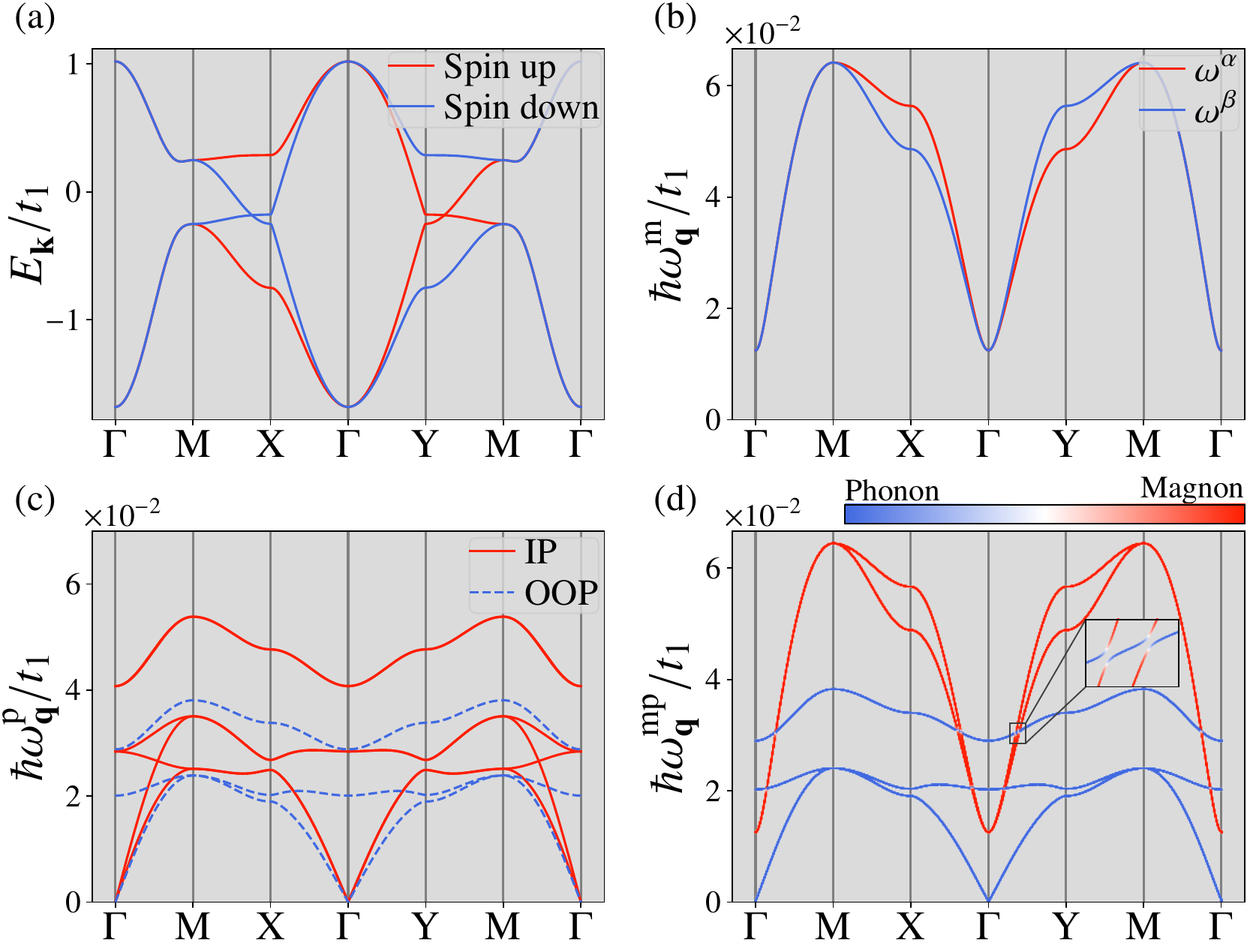}
    \caption{The bands of the electrons and the three different bosons we are considering along a path in the Brillouin zone between high-symmetry points, where numerical parameters are given in Tab.~\ref{tab:Parameters}. (a) The electron bands, (b) the magnon bands, (c) the phonon bands, and (d) the magnon-phonon hybridized bands. The colors in (d) give how magnon- and phonon-like the magnetoelastic modes are [see \eqref{eq:Def of c_n}], where blue means completely phonon-like and red is completely magnon-like.}
    \label{fig:Bands}
\end{figure}

\subsection{Electrons}
\label{sec:Electrons}
We model the electron behavior to include hopping between lattice sites with up to third nearest-neighbor (NN) hopping as well as spin interactions with the magnetic lattice sites.
This model is based on the one utilized in Refs. \cite{BBS23, LMS25, MBS24},
\begin{equation}
\label{eq:Electron Hamiltonian}
\begin{aligned}
    H_\mathrm{el} = &-\sum_{ ij \sigma}t_{ij}c_{i\sigma}^\dag c_{j\sigma}
    -J_{\mathrm{sd}}\sum_{i\sigma\sigma'}\textbf{S}_i\cdot c_{i\sigma}^\dag\boldsymbol{\sigma}_{\sigma\sigma'}c_{i\sigma'}\\-&\mu\sum_{i\sigma}c_{i\sigma}^\dag c_{i\sigma}+\varepsilon_\mr{nm}\sum_{i\in\mr{nm}\sigma}c_{i\sigma}^\dag c_{i\sigma},
\end{aligned}
\end{equation}
where $c_{i\sigma}^{(\dag)}$ annihilates (creates) an electron at site \textit{i} with spin $\sigma$, $\mu$ is the chemical potential, and $\varepsilon_\mr{nm}$ is the on-site energy of the non-magnetic atoms. We have allowed for up to third nearest-neighbor hopping as illustrated in Fig. \ref{fig:Lieb lattice}. For third nearest-neighbor hopping, there are three distinct cases: (i) hopping between two non-magnetic sites with amplitude $t_3$, (ii) hopping between two magnetic sites separated by vacuum with amplitude $t_3^\mathrm{d}$, and (iii) hopping between magnetic sites with a non-magnetic site in between with amplitude $t_3^\mathrm{nm}$. We set these three hoppings equal and use $t_3$ to denote the third neighbor hopping. The second term in \eqref{eq:Electron Hamiltonian} is an on-site interaction between the spins of the itinerant electrons, $\boldsymbol{\sigma}_{\sigma\sigma'}$, and the on-site spins, $\textbf{S}_i$, with coupling strength $J_{\mathrm{sd}}$. 
In several of the candidate Lieb lattice altermagnets, the nonmagnetic site is oxygen. First principle calculations for La$_2$O$_3$Mn$_2$Se$_2$ show that the non-magnetic site is completely filled \cite{GRSMV25}. To match this case, the on-site energy of the non-magnetic site should be sufficiently negative to move one of the bands for each electron spin far below the Fermi level, leaving us with two relevant spin-split bands in ~\figref{fig:Bands}{a}. 

We find the band structure of the electrons by performing a Fourier transformation (FT) and introduce electron band operators $d^\dag_{\textbf{k}\sigma \eta}=\sum_\alpha q_{\textbf{k}\sigma\alpha \eta}c_{\textbf{k}\sigma\alpha}$, where $\eta$ labels the band, $\alpha$ denotes the three sublattices, and $q_{\textbf{k}\sigma\alpha \eta}$ is element $\alpha$ in the eigenvector corresponding to band $\eta$ of the Hamiltonian for an electron with momentum $\textbf{k}$ and spin $\sigma$.
With these new operators, the diagonal Hamiltonian is given by \cite{BBS23}
\begin{equation}
    \label{eq:Diagonal electron Hamiltonian}
    H_\mathrm{el} = \sum_{\textbf{k}\sigma\eta}E_{\textbf{k}\sigma \eta}d^\dag_{\textbf{k}\sigma \eta}d_{\textbf{k}\sigma \eta},
\end{equation}
where $E_{\textbf{k}\sigma \eta}$ is the energy of an electron in band $\eta$ with momentum $\textbf{k}$, and spin $\sigma$. The bands demonstrate a $d$-wave spin splitting, as seen in \figref{fig:Bands}{a}. The Dirac crossings on the line MX for spin up and YM for spin down can be a playground for topological phases in altermagnets \cite{Antonenko2025Lieb, DelRe2025Dirac}. In our case, we keep the Fermi level away from these band crossings and consider the limit of negligible spin-orbit coupling.

\subsection{Bosons}
When calculating the spectral broadening and lifetimes of the electrons in the altermagnet, we will consider the interactions between electrons and either phonons or magnons. As shown in Ref.~\cite{LMS25}, only the IP phonons couple to electrons when considering first order displacements of the lattice. 
In addition to the phonons, the altermagnetic Lieb lattice also hosts magnons due to the broken $SO(3)$-symmetry. Furthermore, these magnons might themselves interact with the OOP phonons \cite{Syljuasen2024} (see Sec.~\ref{sec:Magnon-Phonon Coupling}), creating new hybridized magnetoelastic collective modes.
We also consider the effect of these modes interacting with the electrons.

\subsubsection{Phonons}
\label{sec:phonons}
The phonon bands are determined by studying the displacements, $\textbf{u}_{i\alpha}$, of the lattice atoms away from their equilibrium positions, $\textbf{R}_{i\alpha}$, where \textit{i} refers to a unit cell and $\alpha$, the different atoms in the basis. Expanding the lattice potential around the equilibrium position, keeping terms up to second order in atomic displacements, we find \cite{BruusFlensberg, Syljuasen2024, LMS25, TKWS20, Maeland2025Sep, Maeland2025Dec}
\begin{equation}
\begin{aligned}
    &V(\textbf{r}_{11},\ldots, \textbf{r}_{NN_b}) =  V(\textbf{R}_{11},\ldots, \textbf{R}_{NN_b})\\
&+\frac{1}{2}\sum_{\ia\mu}\sum_{\jb\nu}\left.\frac{\partial^2V}{\partial r_{\ia}^\mu r_{\jb}^\nu}\right|_{\mathrm{eq.}}u^\mu_{\ia}u^\nu_{\jb},
\end{aligned}
\end{equation}
where $\textbf{r}_{\ia}=\textbf{R}_{\ia}+\textbf{u}_{\ia}$, is the instantaneous position of atom $\ia$, \textit{N} is the number of unit cells, $N_b$, the number of atoms in the basis, and $\mu,\nu\in\{x,y,z\}$ denote spatial directions. 
The notation $|_{\mathrm{eq.}}$ indicates evaluation at equilibrium. 
The first term is a constant, which can be disregarded. We have dropped the linear term, since it represents a net force on each atom 
that is zero in equilibrium.
The second-order term can be expressed in terms of the force constant matrix (FCM), defined as
\begin{equation}
    \Phi^{\alpha\beta}_{\mu\nu}(\textbf{R}_i-\textbf{R}_j)\equiv\left.\dfrac{\partial^2V}{\partial r_{\ia}^\mu\partial r_{\jb}^\nu}\right|_{\mathrm{eq.}}.
\end{equation}
The FCM satisfies the general conditions \cite{Lax, Syljuasen2024}
\begin{align}
    \label{eq:FCM self}
    &\sum_{\jb}\Phi_{\mu\nu}^{\alpha\beta}\left(\textbf{R}_i-\textbf{R}_j\right)=0,\\
    \label{eq:FCM symmetries}
    &\Phi_{\mu\nu}^{\alpha\beta}\left(\textbf{R}_i-\textbf{R}_j\right)=\sum_{\mu'\nu'}S^{\raisebox{0.3ex}{$\scriptstyle T$}}_{\mu\mu'}\Phi_{\mu'\nu'}\left(\mathcal{S}\left(\textbf{R}_{\ia}\right)-\mathcal{S}\left(\textbf{R}_{\jb}\right)\right)S_{\nu'\nu},
\end{align}
where $\mathcal{S}$ is a lattice symmetry operation and $S$ is its matrix representation. 
The symmetries of the Lieb lattice are given in \eqref{eq:Lattice symmetries} and provide limitations on the number of free force constants. Following Ref.~\cite{LMS25}, we set all force constants originating from third NNs equal to $\eta$, from second NNs equal to $\rho=\sqrt{2}\eta$, and from first NNs $\gamma=2\eta$. We expect the OOP vibrations to be softer modes than IP, so we set the OOP force constant values to half of their IP counterparts. The numerical values can be found in Tab.~\ref{tab:Parameters}.

The dynamical matrix is defined as the Fourier transform (FT) of the FCM divided by a mass factor, and satisfies an eigenvalue equation \cite{Lax, BruusFlensberg, Klogetvedt2023, Syljuasen2024}, 
\begin{align}
&D^{\alpha\beta}_{\mu\nu}(\textbf{q})\equiv\dfrac{\sum_i\Phi^{\alpha\beta}_{\mu\nu}(\textbf{R}_i-\textbf{R}_j)e^{-i\textbf{q}\cdot(\textbf{R}_{\ia}-\textbf{R}_{\jb})}}{\sqrt{M_\alpha M_\beta}},\\
    \label{eq:Eigenvalue of dynamical matrix}
    \sum_\beta &D^{\alpha\beta}(\textbf{q})\Hat{\boldsymbol{e}}_{\textbf{q}\lambda}^\beta=\left({\omega_{\textbf{q}\lambda}^\mr{p}}\right)^2\Hat{\boldsymbol{e}}_{\textbf{q}\lambda}^{\raisebox{0.5ex}{$\scriptstyle\alpha$}},
\end{align}
where $\Hat{\boldsymbol{e}}_{\textbf{q}\lambda}^\beta$ is the normalized FT of the displacement vector corresponding to phonon mode $\lambda$ and atom $\beta$ within the unit cell. $M_\alpha$ denotes the mass of atom $\alpha$, and $\omega_{\textbf{q}\lambda}^\mr{p}$ is the phonon energy for mode $\lambda$ with momentum $\textbf{q}$.

The displacements are quantized as \cite{Lax, BruusFlensberg}
\begin{equation}
    \label{eq:Displacement phonon}
    \textbf{u}_{i\alpha} = \sum_{\textbf{q}\lambda}\sqrt{\frac{1}{2M_\alpha N\omega^\mr{p}_{\textbf{q}\lambda}}}\Hat{\boldsymbol{e}}_{\textbf{q}\lambda}^{\raisebox{0.5ex}{$\scriptstyle\alpha$}}\left(a_{-\textbf{q},\lambda}^\dag + a_{\textbf{q}\lambda}\right)e^{-i \textbf{R}_{i\alpha}\cdot\textbf{q}},
\end{equation}
where $a_{\textbf{q}\lambda}^{(\dag)}$ annihilates (creates) a phonon with momentum $\textbf{q}$ in mode $\lambda$, and
$N$ is the number of unit cells. Then, the phonon Hamiltonian is
\begin{equation}
    H_{\mr{p}} = \sum_{\mathbf{q}\lambda} \omega_{\textbf{q}\lambda}^\mr{p} a_{\textbf{q},\lambda}^\dag a_{\textbf{q}\lambda}.
\end{equation}
\figref{fig:Bands}{c} shows the phonon spectrum including IP and OOP modes. As expected, there are three acoustic modes and six optical modes when including OOP ion deviations. The highest energy IP optical mode is doubly degenerate with the chosen force constants.

\subsubsection{Magnons}
To derive the magnon properties of the Lieb lattice shown in Fig.~\ref{fig:Lieb lattice}, we consider interactions between the on-site spins of the lattice ions. These occur in three distinct ways: between spins on different sublattices (diagonally), between spins on the same sublattice separated by either vacuum, or a non-magnetic site. We label the sublattice with spin-up \textit{A}, and the spin-down sublattice \textit{B}, such that the Hamiltonian describing these interactions is given by \cite{BBS23, MBS24}
\begin{align}
        \notag
        H_\mathrm{m} =& J_\mathrm{AB}\sum_{\langle i,j\rangle}\textbf{S}_i^A\cdot\textbf{S}_j^B+\sum_{\langle i_x,j_x\rangle}\left(J_\mathrm{nm}\textbf{S}^A_i\cdot\textbf{S}^A_j+J_\mathrm{d}\textbf{S}^B_i\cdot\textbf{S}^B_j\right)\\
    \label{eq:Magnon Hamiltonian}
        &+\sum_{\langle i_y,j_y\rangle}\left(J_\mathrm{d}\textbf{S}^A_i\cdot\textbf{S}^A_j+J_\mathrm{nm}\textbf{S}^B_i\cdot\textbf{S}^B_j\right)\\
        \notag
        &-K_z\sum_i\left(S_{z,i}^AS_{z,i}^A+S_{z,i}^BS_{z,i}^B\right),
\end{align}
where $J_\mathrm{AB}$ denotes the exchange strength between spins on different sublattices, whereas $J_\mathrm{nm}$ and $J_\mathrm{d}$ denote the exchange strength between spins on the same sublattice, either separated by a non-magnetic site or vacuum, respectively. We also include an easy-axis anisotropy in the \textit{z}-direction, whose strength is given by $K_z$. $\textbf{S}_i^{A(B)}$ is the spin operator at site \textit{i} on sublattice \textit{A}(\textit{B}), $\langle i_{x(y)}, j_{x(y)}\rangle$ indicates the nearest-neighbor pairs within the same sublattice along the \textit{x}-(\textit{y}-) direction, and $\langle i,j\rangle$ denotes the nearest-neighbors on different magnetic sublattices.

To diagonalize this Hamiltonian, we follow Ref. \cite{BBS23} and apply a Holstein-Primakoff (HP) transformation combined with a Fourier transformation. The spin fluctuations on the two magnetic sublattices are expressed in terms of the magnon operators $a_\textbf{q}$ and $b_\textbf{q}$, corresponding to sublattices \textit{A} and \textit{B}, respectively. This yields a Hamiltonian that is diagonalized by a Bogoliubov transformation. The result is given by
\begin{align}
    \label{eq:Bogoliubov}
    \begin{pmatrix}
        a_\textbf{q}\\[0.5ex]
        b_{-\textbf{q}}^\dag
    \end{pmatrix}&=\begin{pmatrix}
        u_\textbf{q}&v_\textbf{q}\\[0.5ex]
        v_\textbf{q}^*&u_\textbf{q}^*
    \end{pmatrix}
    \begin{pmatrix}
        \alpha_\textbf{q}\\[0.5ex]
        \beta^\dag_{-\textbf{q}}
    \end{pmatrix},\\[1ex]
    \label{eq:Diagonal magnon Hamiltonian}H_\mathrm{m}&=\sum_\textbf{q}\left(\omega_\textbf{q}^{\mr{m}\alpha}\alpha_\textbf{q}^\dag\alpha_\textbf{q}+\omega_\textbf{q}^{\mr{m}\beta}\beta_\textbf{q}^\dag\beta_\textbf{q}\right),
\end{align}
where $\alpha_\textbf{q}$ and $\beta_\textbf{q}$ are diagonal magnon annihilation operators, and $\omega_\textbf{q}^{\mr{m}\alpha(\beta)}$ is the energy of a magnon with momentum $\textbf{q}$ in mode $\alpha(\beta)$. 
By defining the quantities
\begin{align}
    \notag
    A_\textbf{q} &= 2S\big[J_\mathrm{nm}\cos(2aq_x)+J_\mathrm{d}\cos(2aq_y)\\
    &+2J_\mathrm{AB}-J_\mathrm{nm}-J_\mathrm{d}+K_\mathrm{z}\big],\\[1.3ex]
    B_\textbf{q} &= 4SJ_\mathrm{AB}\cos(aq_x)\cos(aq_y),\\[1.3ex]
    \notag
    C_\textbf{q} &= 2S\big[J_\mathrm{nm}\cos(2aq_y)+J_\mathrm{d}\cos(2aq_x)\\
    &+2J_\mathrm{AB}-J_\mathrm{nm}-J_\mathrm{d}+K_\mathrm{z}\big],
\end{align}
we can express $u_\textbf{q}$, $v_\textbf{q}$, $\omega_\textbf{q}^{\mr{m}\alpha}$, and $\omega_\textbf{q}^{\mr{m}\beta}$ as \cite{BBS23}
\begin{align}
    u_\textbf{q}&=\frac{i}{\sqrt{2}}\sqrt{\frac{A_\textbf{q}+C_\textbf{q}}{\sqrt{\left(A_\textbf{q}+C_\textbf{q}\right)^2-4{B_\textbf{q}\mspace{-3.5mu}}^2}}+1},\\[1.3ex]
    v_\textbf{q}&=\frac{i}{\sqrt{2}}\sqrt{\frac{A_\textbf{q}+C_\textbf{q}}{\sqrt{\left(A_\textbf{q}+C_\textbf{q}\right)^2-4{B_\textbf{q}\mspace{-3.5mu}}^2}}-1},\\[1.3ex]
    \omega_\textbf{q}^{\mr{m}\alpha}&=\dfrac{A_\textbf{q}-C_\textbf{q}}{2}+\dfrac{1}{2}\sqrt{\left(A_\textbf{q}+C_\textbf{q}\right)^2-4{B_\textbf{q}\mspace{-3.5mu}}^2},\\[1.3ex]
    \omega_\textbf{q}^{\mr{m}\beta}&=\dfrac{C_\textbf{q}-A_\textbf{q}}{2}+\dfrac{1}{2}\sqrt{\left(A_\textbf{q}+C_\textbf{q}\right)^2-4{B_\textbf{q}\mspace{-3.5mu}}^2}.
\end{align}

The magnon spectrum is illustrated in ~\figref{fig:Bands}{b}, showing a $d$-wave spin splitting. The Lieb lattice altermagnets have an advantage in that the spin splitting is due to exchange anisotropy already in the next-nearest neighbor exchange ($J_\mathrm{nm} \neq J_\mathrm{d}$) \cite{Chang2025Lieb}. The easy-axis anisotropy breaks a continuous spin rotation symmetry and introduces a gap in the magnon spectrum.

\subsubsection{Magnon-phonon hybridization}
\label{sec:Magnon-Phonon Coupling}
The movements of the ions can lead to tilting of their spin axis, causing the spins to interact with lattice vibrations \cite{Kittel49, Kittel58}, resulting in a coupling between phonons and magnons. For collinear magnets, this type of coupling can be the dominant magnon-phonon interaction \cite{RK14}. By analyzing the anisotropic energy density for a cubic crystal, this interaction takes the form \cite{Thingstad2019magneto, KQ23}
\begin{equation}
\label{eq:Magnon-phonon Hamiltonian}
    H_\mathrm{mp} = \sum_{\langle i,j\rangle}\dfrac{\kappa}{2|\textbf{R}_{ij}|}\dfrac{S_i^{\raisebox{0.5ex}{$\scriptstyle z$}}}{S}\left(\textbf{S}_i\cdot \Hat{\textbf{R}}_{ij}\right)(u_i^z-u_j^z),
\end{equation}
where the notation $\langle i,j\rangle$ indicates a sum over pair of magnetic nearest-neighbors, $|\textbf{R}_{ij}|$ is the distance between atoms \textit{i} and \textit{j} with $\Hat{\textbf{R}}_{ij}$ the corresponding unit vector from \textit{i} to \textit{j}. $\textbf{S}_i$ denotes the spin operator at site \textit{i}, whose normalized \textit{z}-projection $S_i^z/S$ takes values of $\pm1$. $u_i^z$ is the displacement of atom \textit{i} from its equilibrium position along the \textit{z}-axis. $\kappa$ is the magnon-phonon coupling strength. Its numerical value is given in Tab.~\ref{tab:Parameters} and corresponds to the largest coupling while still containing linear phonon-like modes around the $\Gamma$-point.
As demonstrated in App.~\ref{app:Magnon-phonon coupling derivation}, by applying a HP transformation, quantizing the displacements, and applying the Bogoliubov transformation, we arrive at
\begin{align}
\label{eq:Diagonal magnon-phonon Hamiltonian}
    H_\mathrm{mp}&=
    \begin{aligned}[t]
    \sum_{\textbf{q}\lambda}\bigg[&\left(u_\textbf{q}\alpha_{-\textbf{q}}+v_\textbf{q}\beta_\textbf{q}^\dag\right)\left(a_{-\textbf{q}\lambda}^\dag+a_{\textbf{q}\lambda}\right)t^A_{\textbf{q}\lambda}\\[-2ex]
    +&\left(v_\textbf{q}\alpha_\textbf{q}^\dag+u_\textbf{q}\beta_{-\textbf{q}}\right)\left(a_{-\textbf{q}\lambda}^\dag+a_{\textbf{q}\lambda}\right)t^B_{\textbf{q}\lambda}+\mathrm{H.c.}\bigg],
    \end{aligned}\\[1ex]
    t^A_{\textbf{q}\lambda}&\equiv
    \begin{aligned}[t]
        -\dfrac{\kappa}{2a}\sqrt{\dfrac{S}{M_B\omega_{\textbf{q}\lambda}}}\left(\Hat{\boldsymbol{e}}_{\textbf{q}\lambda}\right)^z_B\bigg[&\cos(aq_x)\sin(aq_y)\\[-2ex]
        +i&\sin(aq_x)\cos(aq_y)\bigg],
        \end{aligned}\\[-1ex]
    t^B_{\textbf{q}\lambda}&\equiv
    \begin{aligned}[t]
        -\dfrac{\kappa}{2a}\sqrt{\dfrac{S}{M_A\omega_{\textbf{q}\lambda}}}\left(\Hat{\boldsymbol{e}}_{\textbf{q}\lambda}\right)^z_A\bigg[&\cos(aq_x)\sin(aq_y)\\[-2ex]
        -i&\sin(aq_x)\cos(aq_y)\bigg],
        \end{aligned}
\end{align}
where $\alpha$ and $\beta$ are the diagonal magnon operators, $a_\lambda$ is the annihilation operator for phonon mode $\lambda$, and $\mr{H.c.}$ stands for hermitian conjugate of all preceding terms. Since only OOP (\textit{z}) displacements enter into the magnon–phonon Hamiltonian, \eqref{eq:Magnon-phonon Hamiltonian}, the phonon-mode sum is restricted to the three OOP modes. $\left(\Hat{\boldsymbol{e}}_{\textbf{q}\lambda}\right)^z_A$ is the \textit{z} component of the dynamical-matrix eigenvector [see \eqref{eq:Eigenvalue of dynamical matrix}] associated with the spin-up ion (site $A$). We see that $t^{A/B}_{\textbf{q}\lambda}$, multiplied by either a factor of $u_{\textbf{q}}$ or $v_{\textbf{q}}$, is the coupling between phonons in mode $\lambda$ and the two different magnon modes. 

Having derived the magnon–phonon coupling, we now examine how the phonons modify the magnon properties through hybridization. These hybridized modes define  new quasiparticles that electrons can scatter off, like with phonons and magnons. The total magnon–phonon Hamiltonian consists of three parts: the pure magnon Hamiltonian $H_\mathrm{m}$ from \eqref{eq:Diagonal magnon Hamiltonian}, the pure phonon Hamiltonian $H_\mathrm{p}=\sum_{\textbf{q}\lambda}\omega_{\textbf{q}\lambda}^\mr{p}a_{\textbf{q}\lambda}^\dag a_{\textbf{q}\lambda}$ with energies given by \eqref{eq:Eigenvalue of dynamical matrix}, and the magnon–phonon interaction $H_\mathrm{mp}$ given by \eqref{eq:Diagonal magnon-phonon Hamiltonian}. To describe this full system, we make the system particle-hole symmetric by introducing the Nambu basis
\begin{align}
    \label{eq:Nambudagger}
\Hat{\boldsymbol{\Phi}}_\textbf{q}^\dag&=\!\begin{pmatrix}
        \alpha_\textbf{q}^\dag&\beta_\textbf{q}^\dag&\alpha_{-\textbf{q}}&\beta_{-\textbf{q}}&a_{1\textbf{q}}^\dag&a_{1,-\textbf{q}}&\!a_{2\textbf{q}}^\dag&a_{2,-\textbf{q}}&a_{3\textbf{q}}^\dag&a_{3,-\textbf{q}}
    \end{pmatrix}\!,\\ 
    \label{eq:Nambu}
\Hat{\boldsymbol{\Phi}}_\textbf{q}&=\!\begin{pmatrix}
\alpha_\textbf{q}&\beta_\textbf{q}&\alpha_{-\textbf{q}}^\dag&\beta_{-\textbf{q}}^\dag&a_{1\textbf{q}}&a_{1,-\textbf{q}}^\dag&\!a_{2\textbf{q}}&a_{2,-\textbf{q}}^\dag&a_{3\textbf{q}}&a_{3,-\textbf{q}}^\dag
    \end{pmatrix}^{\!T}\!\!\!\!,
\end{align}
where we have used a hat to indicate that $\Hat{\boldsymbol{\Phi}}$ is structured to simplify the Hamiltonian.
In the Nambu basis, the total magnon-phonon Hamiltonian is given by
\begin{align}
\label{eq:Simple H}
    \mathcal{H}&=\dfrac{1}{2}\sum_\textbf{q}\Hat{\boldsymbol{\Phi}}_\textbf{q}^\dag H(\textbf{q})\Hat{\boldsymbol{\Phi}}_\textbf{q},\\
    H(\textbf{q})&=\begin{pmatrix}
        H_\mathrm{m}(\textbf{q})&\Delta(\textbf{q})\\[1.3ex]
        \Delta^\dag(\textbf{q})&H_\mathrm{p}(\textbf{q})
    \end{pmatrix},
\end{align}
where $H(\textbf{q})$ is a $10\times10$ matrix with
\begin{align}
    H_\mathrm{m}(\textbf{q})&=\mathrm{diag}\left(\omega^{\mr{m}\alpha}_\textbf{q}, \omega^{\mr{m}\beta}_\textbf{q}, \omega^{\mr{m}\alpha}_{-\textbf{q}},\omega^{\mr{m}\beta}_{-\textbf{q}}\right),\\[1ex]
    H_\mathrm{p}(\textbf{q})&=\mathrm{diag}\left(\omega^\mr{p}_{1\textbf{q}}, \omega^\mr{p}_{1,-\textbf{q}}, \omega^\mr{p}_{2\textbf{q}}, \omega^\mr{p}_{2,-\textbf{q}},\omega^\mr{p}_{3\textbf{q}}, \omega^\mr{p}_{3,-\textbf{q}}\right),\\
    \Delta(\textbf{q})&=\begin{pmatrix}
    \textbf{T}^{\raisebox{0.5ex}{$\scriptstyle\alpha$}}_{\textbf{q}}&\textbf{T}^\beta_{\textbf{q}}&\left(\textbf{T}^{\raisebox{0.5ex}{$\scriptstyle\alpha$}}_{-\textbf{q}}\right)^*&\left(\textbf{T}^\beta_{-\textbf{q}}\right)^*
\end{pmatrix}^T,\\[1ex]
    \textbf{T}^{\raisebox{0.5ex}{$\scriptstyle\alpha$}}_{\textbf{q}}&=
    u_\textbf{q}^*\left(\textbf{t}^A_{-\textbf{q}}\right)^*+v_\textbf{q}\textbf{t}_{\textbf{q}}^B,\\[1ex]
    \textbf{T}^\beta_{\textbf{q}}&=
    v_\textbf{q}\textbf{t}^A_{\textbf{q}}+u_\textbf{q}^*\left(\textbf{t}^B_{-\textbf{q}}\right)^*,\\[1ex]
    \textbf{t}^A_{\textbf{q}}&=\begin{pmatrix}
        t^A_{\textbf{q}1}&t^A_{\textbf{q}1}&t^A_{\textbf{q}2}&t^A_{\textbf{q}2}&t^A_{\textbf{q}3}&t^A_{\textbf{q}3}
    \end{pmatrix}^T,\\[1ex]
    \textbf{t}^B_{\textbf{q}}&=\begin{pmatrix}
        t^B_{\textbf{q}1}&t^B_{\textbf{q}1}&t^B_{\textbf{q}2}&t^B_{\textbf{q}2}&t^B_{\textbf{q}3}&t^B_{\textbf{q}3}
    \end{pmatrix}^T.
\end{align}
In App.~\ref{app:Diagonalization of quadratic bosonic Hamiltonian}, we describe the generalized Bogoliubov transformation used to diagonalize this quadratic bosonic Hamiltonian. The diagonalized Hamiltonian for magnon-phonon hybrids is given by
\begin{equation}
    \mathcal{H} = \sum_{\mathbf{q}}\sum_{\lambda=1}^5 \omega_{\mathbf{q}\lambda}^{\text{mp}} \psi_{\lambda\mathbf{q}}^\dagger  \psi_{\lambda\mathbf{q}},
\end{equation}
where $\psi_{\lambda\mathbf{q}}^\dagger$ creates a boson in the magnetoelastic mode $\lambda$ with energy $\omega_{\mathbf{q}\lambda}^{\text{mp}}$.

\label{par:Hybrids couple as magnons}The OOP phonons do not couple directly to electrons to first-order in lattice displacements \cite{LMS25}. Consequently, the magnetoelastic modes interact with electrons as modified magnons. Hence, it is of interest to know how \textit{magnon-like} the magnetoelastic collective modes are. 
To quantify this, we introduce the magnon weight function
\begin{equation}
    \label{eq:Def of c_n}
    \begin{aligned}
        c_\lambda(\textbf{q})\equiv g_{\lambda\lambda}\Big(&\left|T_{1\lambda}(\textbf{q})\right|^2+\left|T_{2\lambda}(\textbf{q})\right|^2\\
        -&\left|T_{6\lambda}(\textbf{q})\right|^2-\left|T_{7\lambda}(\textbf{q})\right|^2\Big),
    \end{aligned}
\end{equation}
where $\lambda$ labels the five magnetoelastic modes, $g\equiv\sigma_z\otimes1_N$ with $\sigma_z$ being the third Pauli matrix and $1_N$ is the $N\times N$ identity matrix, and $T(\textbf{q})$ is the transformation matrix defined by $\boldsymbol{\Phi}=T\boldsymbol{\psi}$, where $\boldsymbol{\psi}$ is a vector consisting of all the creation and annihilation operators for the magnetoelastic modes (see App.~\ref{app:Diagonalization of quadratic bosonic Hamiltonian}). By construction, $c_\lambda\in[0,1]$ with $c_\lambda=1$ corresponding to a purely magnon-like excitation, and $c_\lambda=0$ to a pure phonon-like excitation. 

\figref{fig:Bands}{d} shows the hybridized bands where the color gives \eqref{eq:Def of c_n}. The effect of the hybridization is to turn the original band-crossings between magnons and OOP phonons into anti-crossings.

\section{Electron-boson couplings}
\label{sec:Couplings}
In the previous section, we explored the properties of electrons, phonons, and magnons, as well as coupled magnon-phonon modes, on the Lieb lattice. In this section, we will study how the bosonic collective modes interact with electrons.

\subsection{Electron-phonon interaction}
The electron-phonon coupling is found by allowing each ion in the unit cell \textit{i} and sublattice $\alpha$ to be displaced by $\textbf{u}_{\ia}$ from its equilibrium position. The hopping of the itinerant electrons stems from the overlap of the on-site wavefunctions between different sites, so when the lattice sites move, the hopping will be affected. To capture this, we Taylor-expand the hopping parameters $t_{ij}$ introduced in \eqref{eq:Electron Hamiltonian}, to first order in the displacements \cite{TKWS20, LMS25, Maeland2025Sep, Maeland2025Dec}. We model the on-site wavefunctions as $s$-wave Gaussians with standard deviations $s_x$, $s_y$, and $s_z$ in their respective spatial directions. 
The details of the coupling derivation can be found in Ref.~\cite{LMS25}, which yields the electron-phonon coupling (EPC)
\begin{align}
    \label{eq:Electron-phonon Hamiltonian}
    &H_\mathrm{ep} = \sum_{\textbf{k}\textbf{q} \sigma\eta\eta'\lambda}g_{\lambda\sigma\sigma}^{(\mr{p})\eta'\eta}\left(\textbf{k}+\textbf{q}, \textbf{k}\right)\left(a_{-\textbf{q}\lambda}^\dag + a_{\textbf{q}\lambda}\right)d_{\textbf{k}+\textbf{q},\sigma\eta'}^\dag d_{\textbf{k}\sigma\eta},\\
    &g_{\lambda\sigma\sigma}^{(\mr{p})\eta'\eta}\left(\textbf{k}+\textbf{q},\textbf{k}\right) = \sum_{j=1}^3g_{\lambda\sigma}^{(\mr{p})\eta'\eta(j)}\left(\textbf{k}+\textbf{q},\textbf{k}\right),\\
        \notag&g_{\lambda\sigma\sigma}^{(\mr{p})\eta'\eta(j)}\left(\textbf{k}+\textbf{q},\textbf{k}\right) =\frac{t_j}{2}\sqrt{\frac{1}{2N\omega^\mr{p}_{\textbf{q}\lambda}}}\sum_{\alpha\beta}q_{\textbf{k}+\textbf{q},\sigma\alpha\eta'}^*q_{\textbf{k}\sigma\beta\eta}\\
        \label{eq:Electron-phonon coupling}&\times\sum_{\boldsymbol{\delta}_j}e^{i\textbf{k}\cdot\boldsymbol{\delta}_j}\left(\frac{\delta_{jx}}{s_x^2 },\frac{\delta_{jy}}{s_y^2 }\right)\cdot\left[\frac{\Hat{\boldsymbol{e}}_{\textbf{q}\lambda}^\beta}{\sqrt{M_\beta}}e^{i\textbf{q}\cdot\boldsymbol{\delta}_j}-\frac{\Hat{\boldsymbol{e}}_{\textbf{q}\lambda}^{\raisebox{0.5ex}{$\scriptstyle\alpha$}}}{\sqrt{M_\alpha}}\right],
\end{align}
where $t_1$, $t_2$, and $t_3$ are the hopping amplitudes between nearest, second-nearest, and third-nearest neighbors, respectively.
The vectors connecting these different neighbors are denoted by $\boldsymbol{\delta}_j$. Note that only the IP phonon modes couple to the electrons to first order in the displacements, so $\lambda$ runs over six modes.

\subsection{Electron-magnon interaction}
As demonstrated in Sec.~\ref{sec:Electrons}, the classical part of the interaction between itinerant electrons and the on-site spins, $-J_{\mathrm{sd}}\sum_{i\sigma\sigma'}\textbf{S}_i\cdot c_{i\sigma}^\dag\boldsymbol{\sigma}_{\sigma\sigma'}c_{i\sigma'}$, leads to spin splitting of the electron bands. The quantum part, described by magnons via the HP transformation, gives rise to electron-magnon interactions.
As we are only going to calculate the single-magnon process shown in Fig.~\ref{fig:Sunset diagrams}, we only consider the single-magnon interaction terms \cite{BBS23, MBS24},
\begin{align}
    \notag
     H_\mathrm{em}=&-J_\mathrm{sd}\sqrt{\frac{2S}{N}}\sum_{\textbf{k}\textbf{q}\eta\eta'}\bigg\{\bigg[\left(\Omega^{A\eta\eta'}_{\textbf{k}+\textbf{q},\textbf{k},\downarrow\uparrow}u_\textbf{q}+\Omega^{B\eta\eta'}_{\textbf{k}+\textbf{q},\textbf{k},\downarrow\uparrow}v_\textbf{q}^*\right)\alpha_\textbf{q}\\
     \notag
     &+\left(\Omega^{A\eta\eta'}_{\textbf{k}+\textbf{q},\textbf{k},\downarrow\uparrow}v_\textbf{q}+\Omega^{B\eta\eta'}_{\textbf{k}+\textbf{q},\textbf{k},\downarrow\uparrow}u_\textbf{q}^*\right)\beta_{-\textbf{q}}^\dag\bigg]d_{\textbf{k}+\textbf{q}\downarrow\eta}^\dag d_{\textbf{k}\uparrow\eta'}\\
    \label{eq:Electron-magnon Hamiltonian}&+\bigg[\left(\Omega^{A\eta\eta'}_{\textbf{k}+\textbf{q},\textbf{k},\uparrow\downarrow}u_\textbf{q}^*+\Omega^{B\eta\eta'}_{\textbf{k}+\textbf{q},\textbf{k},\uparrow\downarrow}v_\textbf{q}\right)\alpha_{-\textbf{q}}^\dag\\
    \notag
    &+\left(\Omega^{A\eta\eta'}_{\textbf{k}+\textbf{q},\textbf{k},\uparrow\downarrow}v_\textbf{q}^*+\Omega^{B\eta\eta'}_{\textbf{k}+\textbf{q},\textbf{k},\uparrow\downarrow}u_\textbf{q}\right)\beta_{\textbf{q}}\bigg]d_{\textbf{k}+\textbf{q}\uparrow\eta}^\dag d_{\textbf{k}\downarrow\eta'}\bigg\}.
\end{align}
Here, we have inserted the diagonal electron operators, \eqref{eq:Diagonal electron Hamiltonian}, as well as the diagonal magnon operators from \eqref{eq:Diagonal magnon Hamiltonian}. We also defined $\Omega^{\alpha\eta'\eta}_{\textbf{k}',\textbf{k},\sigma'\sigma}\equiv q^*_{\textbf{k}'\sigma'\alpha\eta'}q_{\textbf{k}\sigma\alpha\eta}$, with $\alpha$ running over the two magnetic sublattices. From \eqref{eq:Electron-magnon Hamiltonian}, we see that the $\alpha$ magnon mode carries spin $S^{\raisebox{0.4ex}{$\scriptstyle z$}}=-1$, while the $\beta$ magnon mode carries spin $S^{\raisebox{0.4ex}{$\scriptstyle z$}}=1$.  
We introduce an index $\chi$ to denote whether the electrons couple
to a magnon creation of destruction operator.
A choice of $\chi=1$ means that a spin-up incoming electron couples to the $\alpha$ magnon, whereas $\chi=-1$ means it couples to $\beta^\dag$. Conversely, for an incoming electron with spin-down, $\chi=1$ corresponds to an interaction with $\beta$, and $\chi=-1$ an interaction with $\alpha^\dag$. 
Expressing \eqref{eq:Electron-magnon Hamiltonian} in a similar form as \eqref{eq:Electron-phonon Hamiltonian}, the interaction takes the form
\begin{widetext}
\begin{align}
    \notag
H_\mathrm{em}=&\sum_{\textbf{k}\textbf{q}\eta\eta'}\bigg\{\bigg[  g^{(\mr{m})\eta\eta'}_{\downarrow\uparrow\chi=1}(\textbf{k}+\textbf{q},\textbf{k})  \alpha_\textbf{q}
     + g^{(\mr{m})\eta\eta'}_{\downarrow\uparrow\chi=-1}(\textbf{k}+\textbf{q},\textbf{k})\beta_{-\textbf{q}}^\dag\bigg]d_{\textbf{k}+\textbf{q}\downarrow\eta}^\dag d_{\textbf{k}\uparrow\eta'} \\
   &+\bigg[ g^{(\mr{m})\eta\eta'}_{\uparrow\downarrow\chi=-1}(\textbf{k}+\textbf{q},\textbf{k})\alpha_{-\textbf{q}}^\dag
    + g^{(\mr{m})\eta\eta'}_{\uparrow\downarrow\chi=1}(\textbf{k}+\textbf{q},\textbf{k})\beta_{\textbf{q}}\bigg]d_{\textbf{k}+\textbf{q}\uparrow\eta}^\dag d_{\textbf{k}\downarrow\eta'}\bigg\}.
  \label{eq:Electron-magnon- Hamiltonian II}   
\end{align}
\end{widetext}
Based on this, we can read off the electron-magnon couplings (EMC) as
\begin{align}
    \label{eq:First electron-magnon coupling}g^{(\mr{m})\eta\eta'}_{\downarrow\uparrow\chi=1}(\textbf{k}+\textbf{q},\textbf{k}) &= p\left(\Omega^{A\eta\eta'}_{\textbf{k}+\textbf{q},\textbf{k},\downarrow\uparrow}u_\textbf{q}+\Omega^{B\eta\eta'}_{\textbf{k}+\textbf{q},\textbf{k},\downarrow\uparrow}v_\textbf{q}^*\right),\\[-2ex]
    g^{(\mr{m})\eta\eta'}_{\downarrow\uparrow\chi=-1}(\textbf{k}+\textbf{q},\textbf{k}) &= p\left(\Omega^{A\eta\eta'}_{\textbf{k}+\textbf{q},\textbf{k},\downarrow\uparrow}v_\textbf{q}+\Omega^{B\eta\eta'}_{\textbf{k}+\textbf{q},\textbf{k},\downarrow\uparrow}u_\textbf{q}^*\right),\\[-2ex]
    g^{(\mr{m})\eta\eta'}_{\uparrow\downarrow\chi=1}(\textbf{k}+\textbf{q},\textbf{k}) &=  p\left(\Omega^{A\eta\eta'}_{\textbf{k}+\textbf{q},\textbf{k},\uparrow\downarrow}v_\textbf{q}^*+\Omega^{B\eta\eta'}_{\textbf{k}+\textbf{q},\textbf{k},\uparrow\downarrow}u_\textbf{q}\right),\\[-2ex]
    \label{eq:Last electron-magnon coupling}
    g^{(\mr{m})\eta\eta'}_{\uparrow\downarrow\chi=-1}(\textbf{k}+\textbf{q},\textbf{k}) &=  p\left(\Omega^{A\eta\eta'}_{\textbf{k}+\textbf{q},\textbf{k},\uparrow\downarrow}u_\textbf{q}^*+\Omega^{B\eta\eta'}_{\textbf{k}+\textbf{q},\textbf{k},\uparrow\downarrow}v_\textbf{q}\right),
\end{align}
where $p\equiv-J_{\mr{sd}}\sqrt{2S/N}$.

\subsection{Electron-magnetoelastic modes interaction}
As discussed in Sec.~\ref{par:Hybrids couple as magnons}, the OOP phonons do not couple to electrons to first order of lattice vibrations, which means that the magnetoelastic modes couple to electrons as magnons modified by the OOP phonons. Because of this, the interaction between the electrons and the magnetoelastic modes will be similar to \eqref{eq:Electron-magnon- Hamiltonian II}, with the magnon operators expressed in terms of the magnetoelastic operators $\boldsymbol{\psi}_\lambda$, introduced in App.~\ref{app:Diagonalization of quadratic bosonic Hamiltonian}.
The structure of the Nambu arrays is such that the first five elements are annihilation operators, and the last five are creation operators. With this structure, we can connect the magnetoelastic mode operators to the magnon- and OOP phonon operators as
\begin{align}
\label{eq:Def transformation matrix}
&\boldsymbol{\Phi}(\textbf{q})=T(\textbf{q})\boldsymbol{\psi}(\textbf{q}),\\
\label{eq:Phi vec}
    &\boldsymbol{\Phi}_\textbf{q}=\!\begin{pmatrix}
        \alpha_\textbf{q}&\beta_\textbf{q}&a_{1\textbf{q}}&\!a_{2\textbf{q}}&a_{3\textbf{q}}&\alpha_{-\textbf{q}}^\dag&\beta_{-\textbf{q}}^\dag&a_{1,-\textbf{q}}^\dag&a_{2,-\textbf{q}}^\dag&a_{3,-\textbf{q}}^\dag
    \end{pmatrix}^{\!T}\!\!\!\!,\\
\label{eq:psi vec}
    &\boldsymbol{\psi}_\textbf{q}=\!\begin{pmatrix}
        \psi_{1\textbf{q}}&\ldots&\psi_{5\textbf{q}}&\psi_{1,-\textbf{q}}^\dag&\ldots&\psi_{5,-\textbf{q}}^\dag
    \end{pmatrix}^{\!T}\!\!\!\!,
\end{align}
where \textit{T} is the transformation matrix (see App.~\ref{app:Diagonalization of quadratic bosonic Hamiltonian}).
With this, we can express the magnon operators in terms of the magnetoelastic operators as
\begin{align}
    \alpha_\textbf{q}&=\sum_{\lambda=1}^5\left[T_{1\lambda}(\textbf{q})\psi_{\lambda\textbf{q}}+T_{1,\lambda+5}(\textbf{q})\psi_{\lambda,-\textbf{q}}^\dag\right],\\
    \beta_\textbf{q}&=\sum_{\lambda=1}^5\left[T_{2\lambda}(\textbf{q})\psi_{\lambda\textbf{q}}+T_{2,\lambda+5}(\textbf{q})\psi_{\lambda,-\textbf{q}}^\dag\right],\\
    \alpha_{-\textbf{q}}^\dag&=\sum_{\lambda=1}^5\left[T_{6\lambda}(\textbf{q})\psi_{\lambda\textbf{q}}+T_{6,\lambda+5}(\textbf{q})\psi_{\lambda,-\textbf{q}}^\dag\right],\\
    \beta_{-\textbf{q}}^\dag&=\sum_{\lambda=1}^5\left[T_{7\lambda}(\textbf{q})\psi_{\lambda\textbf{q}}+T_{7,\lambda+5}(\textbf{q})\psi_{\lambda,-\textbf{q}}^\dag\right].
\end{align}
Inserting these relations into the electron-magnon interaction Hamiltonian, \eqref{eq:Electron-magnon- Hamiltonian II}, yields
\begin{widetext}
\begin{equation}
\label{eq:Electron-hybrid Hamiltonian}
    \begin{aligned}
        H_\mr{emp} = \sum_{\textbf{k}\textbf{q}\eta\eta'\lambda}\bigg\{ \bigg[ &g^{(\mr{mp})\eta\eta'}_{\lambda\downarrow\uparrow\chi=1}(\textbf{k}+\textbf{q},\textbf{k})
        \psi_{\lambda\textbf{q}}+g^{(\mr{mp})\eta\eta'}_{\lambda\downarrow\uparrow\chi=-1}(\textbf{k}+\textbf{q},\textbf{k})
        \psi^\dag_{\lambda,-\textbf{q}}\bigg]d_{\textbf{k}+\textbf{q}\downarrow\eta}^\dag d_{\textbf{k}\uparrow\eta'}\\
        &g^{(\mr{mp})\eta\eta'}_{\lambda\uparrow\downarrow\chi=1}(\textbf{k}+\textbf{q},\textbf{k})
        \psi_{\lambda\textbf{q}}+g^{(\mr{mp})\eta\eta'}_{\lambda\uparrow\downarrow\chi=-1}(\textbf{k}+\textbf{q},\textbf{k})
        \psi^\dag_{\lambda,-\textbf{q}}\bigg]d_{\textbf{k}+\textbf{q}\uparrow\eta}^\dag d_{\textbf{k}\downarrow\eta'}
        \bigg\},
    \end{aligned}
\end{equation}
where, the couplings are given by
\begin{align}
\label{eq:First electron-hybrid coupling}
    g^{(\mr{mp})\eta\eta'}_{\lambda\downarrow\uparrow\chi=1}(\textbf{k}+\textbf{q},\textbf{k})&=g^{(\mr{m})\eta\eta'}_{\downarrow\uparrow\chi=1}(\textbf{k}+\textbf{q},\textbf{k})T_{1\lambda}(\textbf{q})+g^{(\mr{m})\eta\eta'}_{\downarrow\uparrow\chi=-1}(\textbf{k}+\textbf{q},\textbf{k})T_{7\lambda}(\textbf{q}),\\
    g^{(\mr{mp})\eta\eta'}_{\lambda\downarrow\uparrow\chi=-1}(\textbf{k}+\textbf{q},\textbf{k})&=g^{(\mr{m})\eta\eta'}_{\downarrow\uparrow\chi=1}(\textbf{k}+\textbf{q},\textbf{k})T_{1,\lambda+5}(\textbf{q})+g^{(\mr{m})\eta\eta'}_{\downarrow\uparrow\chi=-1}(\textbf{k}+\textbf{q},\textbf{k})T_{7,\lambda+5}(\textbf{q}),\\
    g^{(\mr{mp})\eta\eta'}_{\lambda\uparrow\downarrow\chi=1}(\textbf{k}+\textbf{q},\textbf{k})&=g^{(\mr{m})\eta\eta'}_{\uparrow\downarrow\chi=1}(\textbf{k}+\textbf{q},\textbf{k})T_{2\lambda}(\textbf{q})+g^{(\mr{m})\eta\eta'}_{\uparrow\downarrow\chi=-1}(\textbf{k}+\textbf{q},\textbf{k})T_{6\lambda}(\textbf{q}),\\
\label{eq:Last electron-hybrid coupling}
    g^{(\mr{mp})\eta\eta'}_{\lambda\uparrow\downarrow\chi=-1}(\textbf{k}+\textbf{q},\textbf{k})&=g^{(\mr{m})\eta\eta'}_{\uparrow\downarrow\chi=1}(\textbf{k}+\textbf{q},\textbf{k})T_{2,\lambda+5}(\textbf{q})+g^{(\mr{m})\eta\eta'}_{\uparrow\downarrow\chi=-1}(\textbf{k}+\textbf{q},\textbf{k})T_{6,\lambda+5}(\textbf{q}),
\end{align}
\end{widetext}
with the electron-magnon couplings given by \eqref{eq:First electron-magnon coupling}-\eqref{eq:Last electron-magnon coupling}.

\section{Self-energy}
\label{sec:Self-energy}
\begin{figure}[ht]
    \centering
    \includegraphics[width=\linewidth]{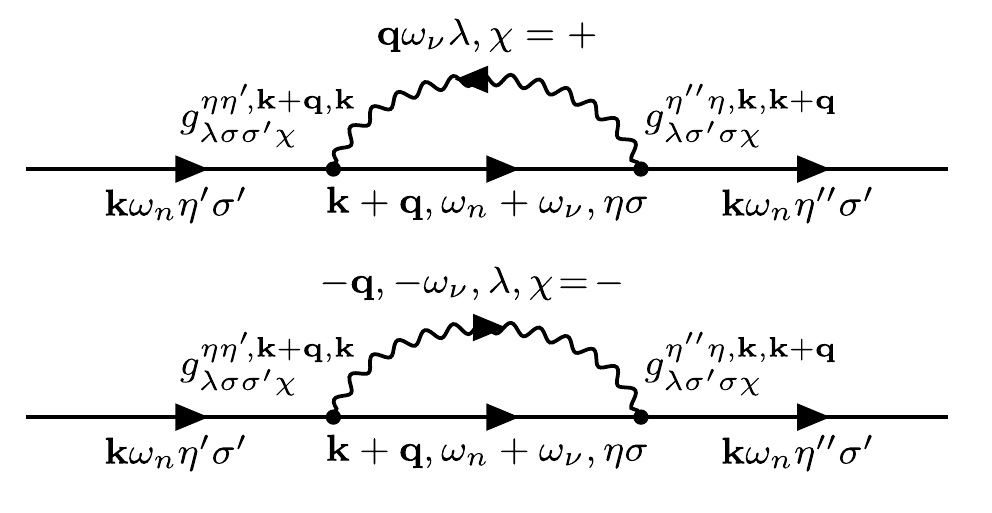}
    \caption{The sunset Feynman diagram, where straight lines represent electrons, and wavy lines represent bosons. The couplings and all relevant quantum numbers are also labeled.}
    \label{fig:Sunset diagrams}
\end{figure}

In the previous section, we derived the interactions between electrons and three different collective bosonic modes. In this section, we compute the renormalization of the electronic band structure due to these interactions.
We truncate the calculation to second order in the electron-boson coupling by employing the Migdal approximation \cite{Migdal58, PGCL07}. The only nonzero second-order Feynman diagram is the sunset diagram illustrated in Fig. \ref{fig:Sunset diagrams}. 
For the various cases where electrons scatter off phonons, magnons, or magnetoelastic collective modes, the electron-boson coupling constant in all cases may be written on the form $g^{\eta\eta'}_{\lambda\sigma\sigma'\chi}(\textbf{k}+\textbf{q},\textbf{k})$. For explicit expressions, see Sec.~\ref{sec:Couplings}. The general self-energy may be compactly expressed in terms of these coupling constants. 
To include finite temperature effects, we utilize the Matsubara formalism of the Green's functions such that the self-energy due to the sunset diagram is given by \cite{MRWS21, BruusFlensberg},
\begin{equation}
    \label{eq:Self-energy first expression}
    \begin{aligned}
        \Sigma_{\sigma'\sigma}^{\eta''\eta'}(\textbf{k},&i\omega_n)=-\sum_{\textbf{q}\eta\lambda\chi}g^{\eta\eta'}_{\lambda\sigma\sigma'\chi}(\textbf{k}+\textbf{q},\textbf{k})g^{\eta''\eta}_{\lambda\sigma'\sigma\chi}(\textbf{k},\textbf{k}+\textbf{q})\\
        &\times k_BT\sum_{\omega_\nu}D^{{\raisebox{0.5ex}{$\scriptstyle \lambda\chi$}}}_0(\textbf{q},i\omega_\nu)G_0^{{\raisebox{0.5ex}{$\scriptstyle \eta\sigma$}}}(\textbf{k}+\textbf{q},i\omega_n+i\omega_\nu),
    \end{aligned}
\end{equation}
\noindent
where $\sigma'$ and $\sigma$ denote the spin of the initial and intermediate electron states. Note that $\sigma$ depends on the boson under consideration. For phonons, $\sigma=\sigma'$, while for magnons or the magnetoelastic modes, $\sigma=-\sigma'$. Further, $\eta$ denotes the electron band, $\lambda$ labels the boson mode, and $\chi$ indicates the boson direction. 
$k_BT$ represents the thermal energy. 
The Matsubara frequencies are defined as $i\omega_n=(2n+1)\pi k_BT$ for fermions and $i\omega_{\nu}=2\pi\nu k_BT$ for bosons. 
The bare boson ($D_0$) and electron ($G_0$) propagators are given by \cite{MRWS21, BruusFlensberg, abrikosov1963methods}
\begin{align}
    G_0^{{\raisebox{0.5ex}{$\scriptstyle \eta\sigma$}}}(\textbf{k}+\textbf{q},i\omega_n)&=\dfrac{1}{i\omega_n-E_{\textbf{k}+\textbf{q,}\eta\sigma}},\\
    D^{{\raisebox{0.5ex}{$\scriptstyle \lambda\chi$}}}_0(\textbf{q},i\omega_\nu)&=\dfrac{\chi}{i\omega_\nu-\chi\omega_{\textbf{q}\lambda}},
\end{align}
where $E_{\textbf{k}+\textbf{q},\eta\sigma}$ is the energy of an electron with spin $\sigma$ and momentum $\textbf{k}+\textbf{q}$ in band $\eta$, while $\omega_{\textbf{q}\lambda}$ is the energy of boson mode $\lambda$ with momentum $\textbf{q}$.
The sum over Matsubara frequencies is done by applying the residue theorem and writing the sum as a complex integral, yielding \cite{mahan, BruusFlensberg}
\begin{equation}
    \begin{aligned}
        &-k_BT\sum_{\omega_\nu}\dfrac{\chi}{i\omega_\nu-\chi\omega_{\textbf{q}\lambda}}\dfrac{1}{i\omega_\nu+i\omega_n-E_{\textbf{k}+\textbf{q},\eta\sigma}}\\
        &=\dfrac{\chi}{i\omega_n-E_{\textbf{k}+\textbf{q},\eta\sigma}+\chi\omega_{\textbf{q}\lambda}}\left[B_\mathrm{E}(\chi\omega_{\textbf{q}\lambda})+F_\mathrm{D}(E_{\textbf{k}+\textbf{q},\eta\sigma})\right],
    \end{aligned}
\end{equation}
where
\begin{align}
    B_\mathrm{E}(\varepsilon)=&\dfrac{1}{e^{\varepsilon/k_BT}-1}=\dfrac{1}{2}\left[\coth(\frac{\varepsilon}{2k_BT})-1\right],\\
    F_\mathrm{D}(\varepsilon)=&\dfrac{1}{e^{\varepsilon/k_BT}+1}=\dfrac{1}{2}\left[1-\tanh(\frac{\varepsilon}{2k_BT})\right],
\end{align}
are the Bose-Einstein and Fermi-Dirac distributions, respectively. We simplify this expression by utilizing the identity $B_\mathrm{E}(-\varepsilon)=-1-B_\mathrm{E}(\varepsilon)$, and the analytic continuation $i\omega_n\rightarrow\omega+i\delta$, where $\delta=0^+$ \cite{BruusFlensberg}, which gives
\begin{equation}
\begin{aligned}
    &-k_BT\sum_{\omega_\nu}D^{{\raisebox{0.5ex}{$\scriptstyle \lambda\chi$}}}_0(\textbf{q}, i\omega_\nu)G_0^{{\raisebox{0.5ex}{$\scriptstyle \eta\sigma$}}}(\textbf{k}+\textbf{q},i\omega_n+i\omega_\nu)\\
    &=\dfrac{1}{2}\dfrac{\coth(\dfrac{\omega_{\textbf{q}\lambda}}{2k_BT})-\chi\tanh(\dfrac{E_{\textbf{k}+\textbf{q},\eta}}{2k_BT})}{\omega-E_{\textbf{k}+\textbf{q},\eta\sigma}+\chi\omega_{\textbf{q}\lambda}+i\delta}.
\end{aligned}
\end{equation}
Inserting this into the full expression of the self-energy, \eqref{eq:Self-energy first expression}, as well as turning the momentum sum into an integral,
\begin{equation}
    \sum_\textbf{q}\rightarrow\dfrac{Na^2}{\pi^2}\int_{-\textstyle\frac{\pi}{2a}}^{\textstyle\frac{\pi}{2a}}\dd q_x\int_{-\textstyle\frac{\pi}{2a}}^{\textstyle\frac{\pi}{2a}}\dd q_y,
\end{equation}
the self-energy is expressed as
\begin{equation}
\label{eq:Self energy complex}
    \begin{aligned}
         &\Sigma_{\sigma'\sigma}^{\eta''\eta'}(\textbf{k},\omega)=\dfrac{Na^2}{2\pi^2}\sum_{\eta\lambda\chi}\int
         _{-\textstyle\frac{\pi}{2a}}^{\textstyle\frac{\pi}{2a}}\!\!\dd q_x\int_{-\textstyle\frac{\pi}{2a}}^{\textstyle\frac{\pi}{2a}}\!\!\dd q_y g_{\lambda\sigma\sigma'\chi}^{\eta\eta'}(\textbf{k}+\textbf{q},\textbf{k}) \\
    &\times g_{\lambda\sigma'\sigma\chi}^{\eta''\eta}(\textbf{k},\textbf{k}+\textbf{q})\dfrac{\coth(\dfrac{\omega_{\textbf{q}\lambda}}{2k_BT})-\chi\tanh(\dfrac{E_{\textbf{k}+\textbf{q},\eta\sigma}}{2k_BT})}{\omega-E_{\textbf{k}+\textbf{q},\eta\sigma}+\chi\omega_{\textbf{q}\lambda}+i\delta}.
    \end{aligned}
\end{equation}
We proceed by noting that the couplings between the electrons and the three bosons are either completely real (electron-phonon) or completely imaginary (electron-magnon and electron-magnetoelastic). Hence, the product of couplings in \eqref{eq:Self energy complex} is purely real. This implies that the imaginary part of $\Sigma_{\sigma'\sigma}^{\eta''\eta'}(\textbf{k},\omega)$ comes from the $i\delta$-term in the denominator.

\subsection{Imaginary part of the self-energy}
Separating \eqref{eq:Self energy complex} into the real and imaginary parts, we see that the imaginary part will be proportional to a Lorentzian with a width determined by $\delta$, which in the limit $\delta \to 0^+$ yields a Dirac delta function 
$\delta(\omega-E_{\textbf{k}+\textbf{q},\eta\sigma}+\chi\omega_{\textbf{q}\lambda})$. We will therefore compute the imaginary part of $\Sigma$ first, and then obtain the real part using the Kramer-Kronig relation. 
Taking the limit $\delta\rightarrow0^+$ of \eqref{eq:Self energy complex}, we obtain
\begin{equation}
\label{eq:Self energy imaginary part first}
    \begin{aligned}
        &\mathrm{Im}\Sigma_{\sigma'\sigma}^{\eta''\eta'}(\textbf{k},\omega)=-\dfrac{Na^2}{2\pi}\sum_{\eta\lambda\chi}\int_{-\textstyle\frac{\pi}{2a}}^{\textstyle\frac{\pi}{2a}}\!\!\dd q_x\int_{-\textstyle\frac{\pi}{2a}}^{\textstyle\frac{\pi}{2a}}\!\!\dd q_y \\
        &\times g_{\lambda\sigma\sigma'\chi}^{\eta\eta'}(\textbf{k}+\textbf{q},\textbf{k})g_{\lambda\sigma'\sigma\chi}^{\eta''\eta}(\textbf{k},\textbf{k}+\textbf{q})\\
        &\times\left[\coth(\dfrac{\omega_{\textbf{q}\lambda}}{2k_BT})-\chi\tanh(\dfrac{E_{\textbf{k}+\textbf{q},\eta\sigma}}{2k_BT})\right]\delta(\xi_{\textbf{q}w}),
    \end{aligned}
\end{equation}
where we defined $\xi_{\textbf{q}w}\equiv\omega-E_{\textbf{k}+\textbf{q},\eta\sigma}+\chi\omega_{\textbf{q}\lambda}$, and $w=\{\omega,\textbf{k},\eta,\sigma,\chi,\lambda\}$. The two-dimensional integral can be expressed as a line integral by utilizing the general relation
\begin{equation}
    \int\!\dd^2qf(\textbf{q})\delta(\xi_{\textbf{q}w})=\int_\zeta\!\dd\textbf{q}\cdot\Hat{\textbf{e}}_\textbf{q}\dfrac{f(\textbf{q})}{\left|\boldsymbol{\nabla}_\textbf{q}\xi_{\textbf{q}w}\right|},
\end{equation}
where $\zeta$ is the curve satisfying $\xi_{\textbf{q}w}=0$, $\Hat{\textbf{e}}_\textbf{q}$ is the unit vector pointing along this curve, and $\boldsymbol{\nabla}_\textbf{q}$ is the nabla operator in $q$-space. Inserting this relation into \eqref{eq:Self energy imaginary part first} yields
\begin{equation}
\label{eq:Self-energy imaginary part}
    \begin{aligned}
        &\mathrm{Im}\Sigma_{\sigma'\sigma}^{\eta''\eta'}(\textbf{k},\omega)=-\dfrac{Na^2}{2\pi}\sum_{\eta\lambda\chi}\int_\zeta\!\dd\textbf{q}\cdot\Hat{\textbf{e}}_\textbf{q}\\
        &\times g_{\lambda\sigma\sigma'\chi}^{\eta\eta'}(\textbf{k}+\textbf{q},\textbf{k})\left[g_{\lambda\sigma\sigma'\chi}^{\eta\eta''}(\textbf{k}+\textbf{q},\textbf{k})\right]^*\\
        &\times\left[\coth(\dfrac{\omega_{\textbf{q}\lambda}}{2k_BT})-\chi\tanh(\dfrac{E_{\textbf{k}+\textbf{q},\eta\sigma}}{2k_BT})\right]\dfrac{1}{\left|\boldsymbol{\nabla}_\textbf{q}\xi_{\textbf{q}w}\right|},
    \end{aligned}
\end{equation}
where $\zeta$ is restricted to the first Brillouin zone (BZ), and we inserted the relation 
\begin{equation}
\label{eq:Coupling hermitian}
    g_{\lambda\sigma'\sigma\chi}^{\eta''\eta}(\textbf{k},\textbf{k}+\textbf{q})=\left[g_{\lambda\sigma\sigma'\chi}^{\eta\eta''}(\textbf{k}+\textbf{q},\textbf{k})\right]^*,
\end{equation}
stemming from the requirement that the interaction Hamiltonians, \eqref{eq:Electron-phonon Hamiltonian}, \eqref{eq:Electron-magnon- Hamiltonian II}, and \eqref{eq:Electron-hybrid Hamiltonian}, are hermitian. 
The electron-boson couplings \eqref{eq:Electron-phonon coupling}, \eqref{eq:First electron-magnon coupling}-\eqref{eq:Last electron-magnon coupling}, and \eqref{eq:First electron-hybrid coupling}-\eqref{eq:Last electron-hybrid coupling} depend explicitly on the eigenvectors of the electronic Hamiltonian. Consequently, the off-diagonal elements in \eqref{eq:Self-energy imaginary part} $(\eta''\neq\eta')$ inherit the associated $U(1)$ phase (gauge) freedom. Nevertheless, \eqref{eq:Self-energy first expression} and \eqref{eq:Coupling hermitian} give that the off-diagonal elements are related by complex conjugation,
\begin{equation}
    \label{eq:Off-diagonal complex conjugate}
    \Sigma^{\eta''\eta'}_{\sigma'\sigma}(\textbf{k},\omega) = \left[\Sigma^{\eta'\eta''}_{\sigma'\sigma}(\textbf{k},\omega)\right]^*\!,\quad \eta''\neq\eta'.
\end{equation} 
As seen from \eqref{eq:pole} and \eqref{eq:Spectral function final expression}, the off-diagonal terms enter physical observables through products with each other, so any dependence on the eigenvector gauge cancels.

We are now able to evaluate \eqref{eq:Self-energy imaginary part} numerically, where the main numerical challenge is accurately locating the $\zeta$ curves. The algorithm used is described in App.~\ref{App:Nodal lines}.

\subsection{Real part of the self-energy}
The real part of the self-energy can be found from the imaginary part by applying the Kramers-Kronig relation \cite{Toll56}
\begin{equation}
\mathrm{Re}\Sigma_{\sigma'\sigma}^{\eta''\eta'}(\textbf{k},\omega')=\frac{1}{\pi}
     P\int_{-\infty}^\infty\!\!\dd\omega\frac{\mathrm{Im}\Sigma_{\sigma'\sigma}^{\eta''\eta'}(\textbf{k},\omega)}{\omega-\omega'},
\end{equation}
where \textit{P} is the Cauchy principal value. Numerically, we calculate this integral with the trapezoidal rule and choose the values of $\omega'$ to lie in the middle of the $\omega$ values used to calculate the imaginary part of the self-energy to avoid the divergence at the point $\omega=\omega'$. 

\subsection{Electron self-energies on the Lieb lattice}
\begin{figure}[tb]
    \centering
    \includegraphics[width=\linewidth]{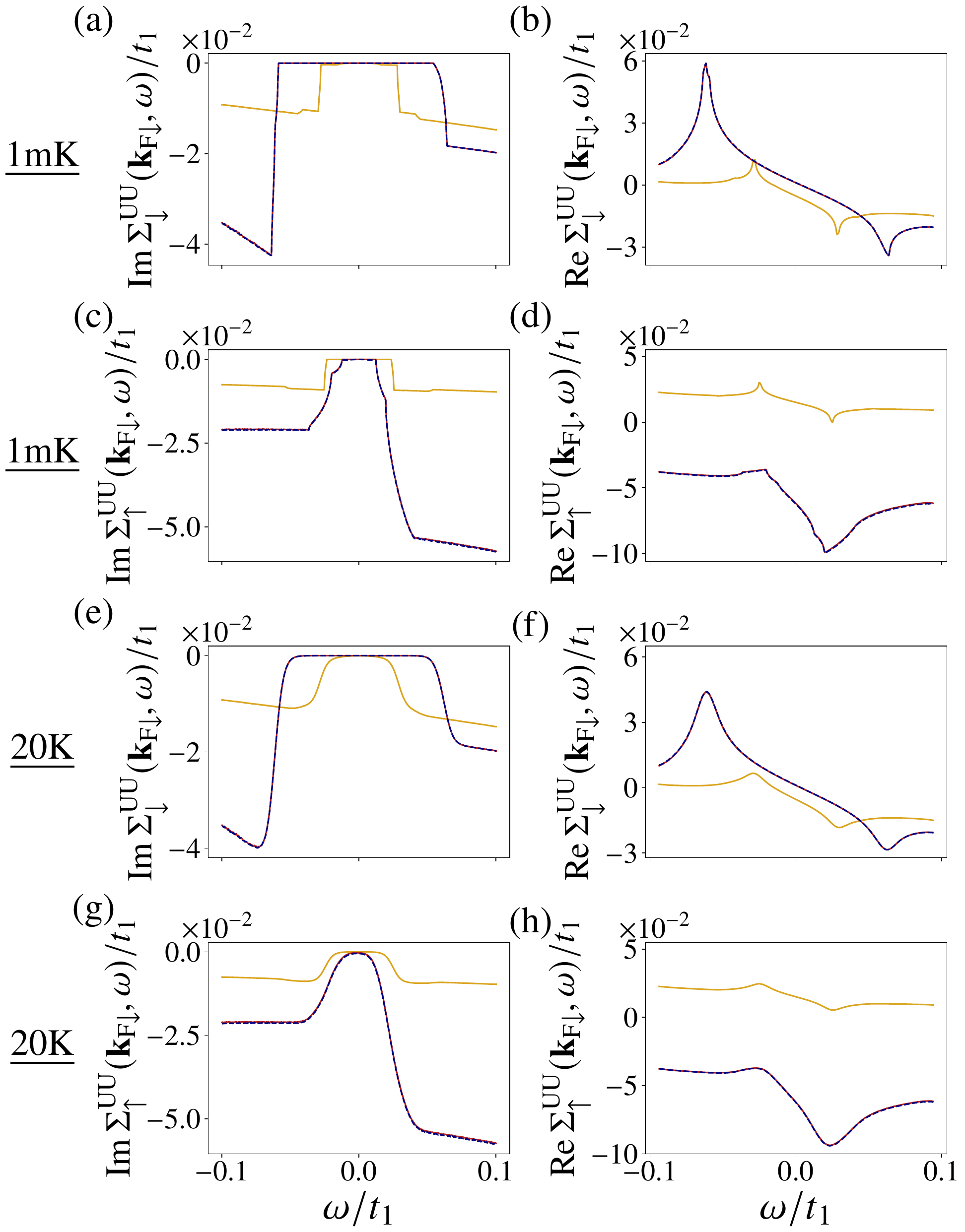}
    \caption{The real and imaginary parts of the electron self-energy for scattering of electrons off three different bosonic modes: phonons (yellow line), magnons (red line), and magnetoelastic modes (blue dashed line). The first and second rows show the zero-temperature limit ($T=1$~mK), whereas the third and fourth rows show the case with $ T=20$~K. 
    The first and third rows show results when the external electron has spin down, while the second and fourth rows show results for electrons with spin up. $\textbf{k}_{\mr{F}\downarrow}$ lies on the FS of spin-down electrons along the X-$\Gamma$ line. The values of the other parameters that are used are given in Tab.~\ref{tab:Parameters}.}
    \label{fig:SE}
\end{figure}

In Fig.~\ref{fig:SE}, we show the imaginary part
$\mathrm{Im}\Sigma_{\sigma'\sigma}^{\eta''\eta'}(\textbf{k},\omega)$ of the electronic self-energy, for all three cases of collective bosonic modes.
When computing  $\Sigma_{\sigma'\sigma}^{\eta''\eta'}(\textbf{k},\omega')$, we consider $\textbf{k}$ to lie on the FS of spin-down electrons along the X-$\Gamma$ line, $\textbf{k} = \textbf{k}_{\mr{F}\downarrow}$. We focus on the case $\eta''=\eta'=U$, where \textit{U} refers to the upper band, as the Fermi level lies in this band. In addition, the off-diagonal elements are negligible as the Fermi level is far from the band-crossings.

Around $\omega=0$, the imaginary part of the electron self-energy, $\mathrm{Im}\Sigma_{\sigma'\sigma}^{\eta''\eta'}(\textbf{k},\omega)$, has a plateau where it is approximately zero for all three bosonic modes. In the case of magnons, this plateau reflects the easy-axis anisotropy gap, requiring a finite threshold energy to create a magnon. The plateau width for the magnon case is spin-dependent, as magnons induce a spin flip of the electrons. Consider an electron with spin down and momentum $\textbf{k}$. When it interacts with a magnon, the virtual state will be an electron with spin up and momentum $\textbf{k}+\textbf{q}$. Since $\textbf{k}$ is not close to the FS of spin-up electrons, the virtual state requires a nonzero $\textbf{q}$ and so a larger magnon energy. In the opposite case of an original spin-up electron, the virtual state will have spin down and momentum $\textbf{k}+\textbf{q}$, which will be close to the FS of spin-down electrons with small $|\textbf{q}|$. Hence, this process requires less energy and can occur at a smaller $|\omega|$ equal to the magnon gap \cite{MRWS21}.

Let us now consider the imaginary part of the electron self-energy when electrons scatter off phonons. As noted above, only in-plane phonons couple directly to electrons to linear order in displacements. There are six in-plane phonon branches, two acoustical and four optical. The acoustic modes are gapless and hence feature zero-energy excitations. 
Under such circumstances, one expects the acoustical phonons to give a non-zero contribution to 
$\mathrm{Im} \Sigma$. However, 
in Fig.~\ref{fig:SE}, also for phonons there appears to be a plateau in $\mathrm{Im} \Sigma$. In reality, there is a small non-zero part of $\mathrm{Im} \Sigma \sim \omega^2$ which is not visible on the scale shown here, originating from the acoustic phonons, which vanishes at $\omega=0$. 
The small value of $\mathrm{Im} \Sigma$ originating with these phonon branches indicates a weak coupling between the electrons and the acoustical phonons. 
Furthermore, for the phonon-case, we see two step-like features in $\mathrm{Im} \Sigma$ in Fig.~\ref{fig:SE}. The first step is due to the two optical phonon branches with the smallest gap. 
The branches feature a flat structure around the $\Gamma$-point, as can be seen in \figref{fig:Bands}{c}. The first step is located at $|\omega|$ corresponding to the energy of these phonon branches, while the magnitude of the first step is due to the flatness of the optical phonon branches around the $\Gamma$-point, giving a large phonon density of states. 
The second step is located at $|\omega|$ corresponding to the energy gap of the two upper degenerate optical phonon bands. 

We next return to a more detailed discussion of $\mathrm{Im} \Sigma$ for the case of magnons (and magnetoelastic modes). 
Namely, there is a distinct asymmetry in the steps in $\mathrm{Im} \Sigma$ for $\omega < 0$ and $\omega >0$. 
This contrasts with the case of phonons,
where one essentially can connect $\mathrm{Im} \Sigma$ for $\omega < 0$ and $\omega > 0$ by extrapolation through the region $|\omega| < \omega_{\text{max}}$, where $\omega_{\text{max}}$ is the maximum boson energy.
The origin of this asymmetry lies in which magnon mode couples to electrons on either side of the plateau, combined with the altermagnetic spin splitting of the magnon modes. 
Studying the electron-magnon interaction given by \eqref{eq:Electron-magnon- Hamiltonian II}, we see that the magnon mode is given by $\chi$ and the spin of the external electron. In App.~\ref{app:Determining magnon species}, we show that the most significant contribution to the imaginary part of the self-energy comes when $\chi$ and $\omega$ have opposite signs. Hence, the magnon mode dominating the contribution to the self-energy changes at $\omega=0$. For example, if the external electron has spin up, the self-energy below $\omega=0$ is dominated by the $\alpha$-mode with $\chi=1$, and above $\omega=0$ the $\beta$-mode with $\chi=-1$ dominates.
This is exactly what we observe in 
\figref{fig:SE}{a} and \figref{fig:SE}{c}, the two spins couple to opposite modes on each side on $\omega=0$. 
The implication of this is that we are able to distinguish between spin-up and spin-down electrons even in non-spin-resolved experimental methods such as regular ARPES (see Sec.~\ref{sec:Spectral function}). 

From Fig.~\ref{fig:SE}, it is seen that the magnon-phonon coupling has very little impact on the electron self-energy. The results are virtually indistinguishable in the cases where electrons scatter off pure magnons and collective magnetoelastic modes.
This is because the effect of the magnon-phonon coupling is to convert the original magnon-phonon band crossings into anti-crossings. Hence, the effect of the hybridization is only noticeable in small regions of the BZ.
However, we emphasize that when determining magnon and phonon properties per se, magnetoelastic effects might be crucial, as in Ref.~\cite{Thingstad2019magneto}. Below, we will consider electron spectral functions. Based on the above, we will consider only the pure phonon and pure magnon cases, noting that all conclusions drawn for the pure magnon case also apply to the case where electrons scatter off collective magnetoelastic modes.

\section{Spectral function}
\label{sec:Spectral function}
To experimentally probe the effects of the various collective modes on the electronic self-energies, angular resolved photoemission spectroscopy 
(ARPES) and spin-resolved ARPES measurements, has proven to be a very useful tool, see for instance \cite{RevModPhys.75.473,RevModPhys.93.025006,Damascelli04}. ARPES yields direct information on the many-body renormalization of electronic energy bands, including broadening and shifts. In these experiments, the measured quantity is an intensity, which is proportional to the spectral function, given by \cite{RevModPhys.75.473,RevModPhys.93.025006,Damascelli04}
\begin{equation}
    \label{eq:Def spectral function}
    A_{\sigma'\sigma}(\textbf{k},\omega) = -\pi^{-1}\mr{Tr}\left[\mr{Im}G_{\sigma'\sigma}(\textbf{k},\omega)\right],
\end{equation}
where $\sigma'$ is the spin of the initial electron and $\sigma$ is the spin of the off-shell electron. $\sigma$ depends on the boson under consideration, for electron-phonon interactions $\sigma=\sigma'$, while for interactions with magnons or magnetoelastic modes $\sigma=-\sigma'$. The Green's function $G(\textbf{k},\sigma)$ is, in general, a $4\times4$ matrix in the two electron band indices and spin. Since the interaction depicted in Fig.~\ref{fig:Sunset diagrams} is diagonal in spin, we consider two spin-resolved Green's functions. In this case, the Green's functions are two independent $2\times2$ matrices. The Dyson equation expresses the full Green's function $G$ in terms of the non-interacting Green's function $G_0$ and the self-energy $\Sigma$ \cite{Dyson1949},
\begin{equation}
\begin{aligned}
\label{eq:Dyson}
&G^{-1}_{\sigma'\sigma}(\textbf{k},\omega) = G_{0,\sigma'}^{-1}(\textbf{k},\omega)-\Sigma_{\sigma'\sigma}(\textbf{k},\omega)\\
&=\begin{pmatrix}\omega-E_{\textbf{k}\mr{U}\sigma'}-\Sigma^{\mr{UU}}_{\sigma'\sigma}(\textbf{k},\omega)&-\Sigma^{\mr{UL}}_{\sigma'\sigma}(\textbf{k},\omega)\\[1.0ex]
-\Sigma^{\mr{LU}}_{\sigma'\sigma}(\textbf{k},\omega)&\omega-E_{\textbf{k}\mr{L}\sigma'}-\Sigma^{\mr{LL}}_{\sigma'\sigma}(\textbf{k},\omega) 
\end{pmatrix},
\end{aligned}
\end{equation}
where U and L label the upper and lower electron bands, respectively. Inverting this matrix, we find
\begin{align}
    \notag
    &G_{\sigma'\sigma}(\textbf{k},\omega) = \dfrac{1}{\Theta_{\sigma'\sigma}(\textbf{k},\omega)}\\
    \label{eq:Expression G}
    &\times\begin{pmatrix}\omega-E_{\textbf{k}\mr{L}\sigma'}-\Sigma^{\mr{LL}}_{\sigma'\sigma}(\textbf{k},\omega)&\Sigma^{\mr{UL}}_{\sigma'\sigma}(\textbf{k},\omega)\\[1.0ex]
\Sigma^{\mr{LU}}_{\sigma'\sigma}(\textbf{k},\omega)& \omega-E_{\textbf{k}\mr{U}\sigma'}-\Sigma^{\mr{UU}}_{\sigma'\sigma}(\textbf{k},\omega)
\end{pmatrix},\\
\notag
&\Theta_{\sigma'\sigma}(\textbf{k},\omega)\equiv\left[\omega-E_{\textbf{k}\mr{U}\sigma'}-\Sigma^{\mr{UU}}_{\sigma'\sigma}(\textbf{k},\omega)\right]\\
&\times\left[\omega-E_{\textbf{k}\mr{L}\sigma'}-\Sigma^{\mr{LL}}_{\sigma'\sigma}(\textbf{k},\omega)\right]-\Sigma^{\mr{UL}}_{\sigma'\sigma}(\textbf{k},\omega)\Sigma^{\mr{LU}}_{\sigma'\sigma}(\textbf{k},\omega),
\end{align}
where $\Theta$ is the determinant of the inverse matrix.
The singularities in the Green's function are related to the renormalized bands and appear at
\begin{align}
    \omega_\pm &= \frac{1}{2}\bigg[ E_{\textbf{k}\mr{U}\sigma'}+\Sigma^{\mr{UU}}_{\sigma'\sigma}+E_{\textbf{k}\mr{L}\sigma'}+\Sigma^{\mr{LL}}_{\sigma'\sigma} \nonumber\\
    &\pm \pqty{\left[E_{\textbf{k}\mr{U}\sigma'}+\Sigma^{\mr{UU}}_{\sigma'\sigma}-E_{\textbf{k}\mr{L}\sigma'}-\Sigma^{\mr{LL}}_{\sigma'\sigma}\right]^2 + 4 \left|\Sigma^{\mr{UL}}_{\sigma'\sigma} \right|^2}^{\frac12} \bigg], \label{eq:pole}
\end{align}
which must be solved self-consistently. We utilized \eqref{eq:Off-diagonal complex conjugate} to get $|\Sigma^{\mr{UL}}_{\sigma'\sigma} |^2$ and omitted the momentum and frequency dependence of the self-energies for notational convenience.
With this Green's function, the spectral function, \eqref{eq:Def spectral function}, becomes
\begin{widetext}
\begin{equation}
\begin{aligned}
    \label{eq:Spectral function final expression}
    A_{\sigma'\sigma}(\textbf{k},\omega)&=-\pi^{-1}\mr{Im}\left(\dfrac{\omega-E_{\textbf{k}\mr{U}\sigma'}-\Sigma^{\mr{UU}}_{\sigma'\sigma}(\textbf{k},\omega)+\omega-E_{\textbf{k}\mr{L}\sigma'}-\Sigma^{\mr{LL}}_{\sigma'\sigma}(\textbf{k},\omega)}{\left[\omega-E_{\textbf{k}\mr{U}\sigma'}-\Sigma^{\mr{UU}}_{\sigma'\sigma}(\textbf{k},\omega)\right]\left[\omega-E_{\textbf{k}\mr{L}\sigma'}-\Sigma^{\mr{LL}}_{\sigma'\sigma}(\textbf{k},\omega)\right]-\left|\Sigma^{\mr{UL}}_{\sigma'\sigma}(\textbf{k},\omega)\right|^2}\right)\\
    &\approx-\frac{1}{\pi}\left(\frac{\mr{Im}\Sigma^{\mr{UU}}_{\sigma'\sigma}(\textbf{k},\omega)}{\left(\omega-E_{\textbf{k}\mr{U}\sigma'}-\mr{Re}\Sigma^{\mr{UU}}_{\sigma'\sigma}(\textbf{k},\omega)\right)^2 + \left(\mr{Im}\Sigma^{\mr{UU}}_{\sigma'\sigma}(\textbf{k},\omega)\right)^2}+\frac{\mr{Im}\Sigma^{\mr{LL}}_{\sigma'\sigma}(\textbf{k},\omega)}{\left(\omega-E_{\textbf{k}\mr{L}\sigma'}-\mr{Re}\Sigma^{\mr{LL}}_{\sigma'\sigma}(\textbf{k},\omega)\right)^2+\left(\mr{Im}\Sigma^{\mr{LL}}_{\sigma'\sigma}(\textbf{k},\omega)\right)^2}\right),
\end{aligned}
\end{equation}
\end{widetext}
where, in the last step we used that unless $E_{\textbf{k}\mr{U}\sigma'} \approx E_{\textbf{k}\mr{L}\sigma'}$, the off-diagonal elements play a small role in the location of the renormalized bands, assuming $|\Sigma^{\mr{UL}}_{\sigma'\sigma}(\textbf{k},\omega)|$ is always small compared to the electron bandwidth.
The physics close to the FS is most important, so since we always choose the FS far from any band crossing, we can neglect off-diagonal elements of the self-energy. 
Note that in all figures we plot the full expression given by the first line in \eqref{eq:Spectral function final expression}.
However, by neglecting the off-diaginal elements $\Sigma^{\mr{UL}}_{\sigma'\sigma}(\textbf{k},\omega)$, it becomes easier to gain insight into the structure of the spectral function. In this case, one recognizes the spectral weight as the sum of two Lorentzians. For the upper band the center of the electron band is determined from $\omega-E_{\textbf{k}\mr{U}\sigma'}-\mr{Re}\Sigma^{\mr{UU}}_{\sigma'\sigma}(\textbf{k},\omega)=0$ while the half-width at half-maximum is given by $\mr{Im}\Sigma^{\mr{UU}}_{\sigma'\sigma}(\textbf{k},\omega)$. Corresponding expressions apply for the lower band. 
To avoid unphysical Lorentzians with zero width, we add a small constant contribution of $-5$~meV to the diagonal elements of the imaginary part of the self-energy \cite{MRWS21}.

\begin{figure}[tb]
    \centering
    \includegraphics[width=\linewidth]{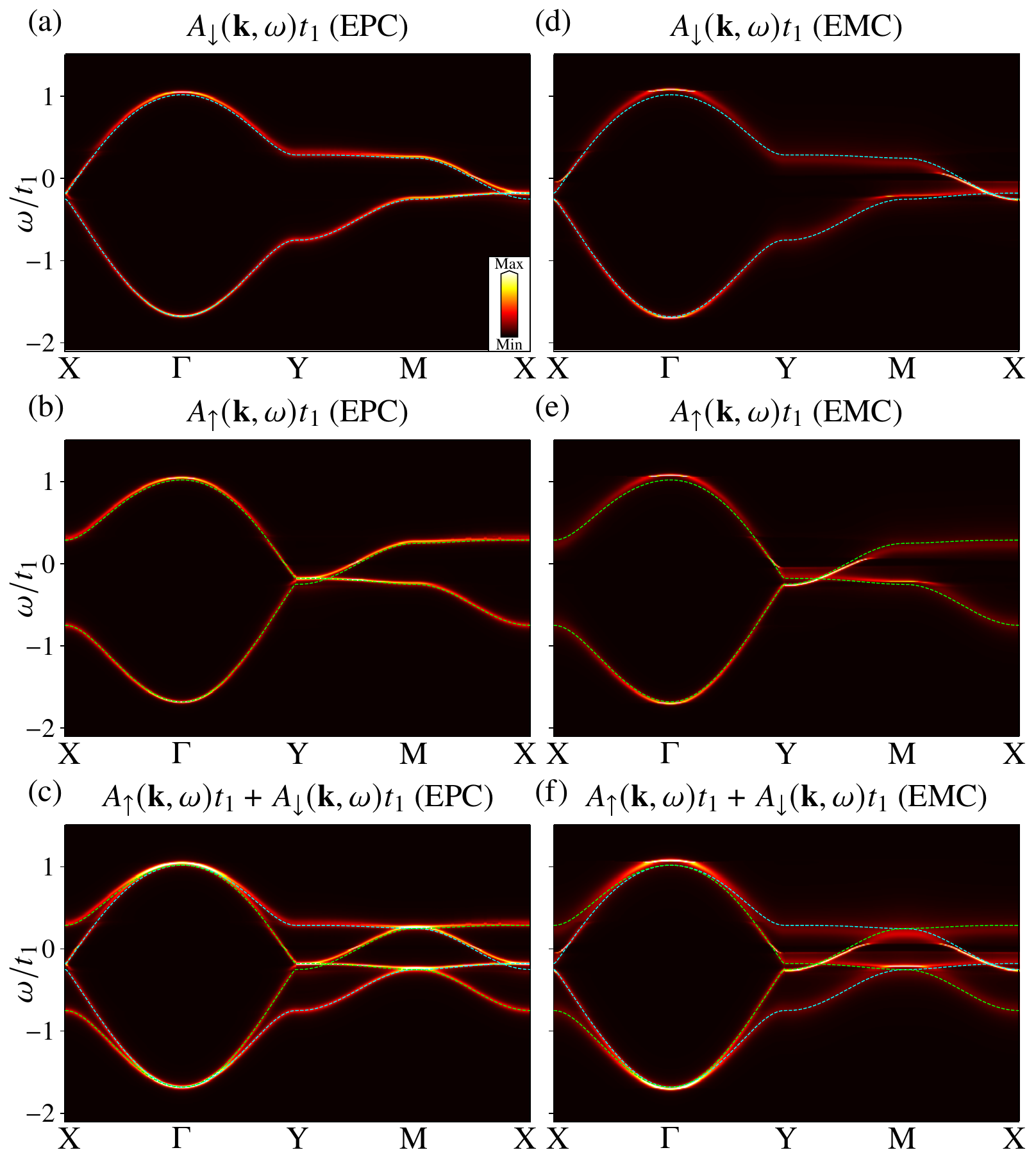}
    \caption{The spectral function as function of both $\textbf{k}$ along the \textit{x}-axis and $\omega$ along the \textit{y}-axis. The values of $\textbf{k}$ follow a path in the BZ between high-symmetry points. The dashed blue (green) lines indicate the original bands for spin-down (up) electrons. We consider the case $T=20$~K in all of the plots and (a) spin-down electrons interacting with phonons, (b) spin-up electrons interacting with phonons, (c) the sum of these, (d) spin-down electrons interacting with magnons, (e) spin-up electrons interacting with magnons, and (f) the sum of these.
    For the rest of the parameters used, see Tab.~\ref{tab:Parameters}.}
    \label{fig:Spectral functions}
\end{figure}

\begin{figure}
    \centering
    \includegraphics[width=\linewidth]{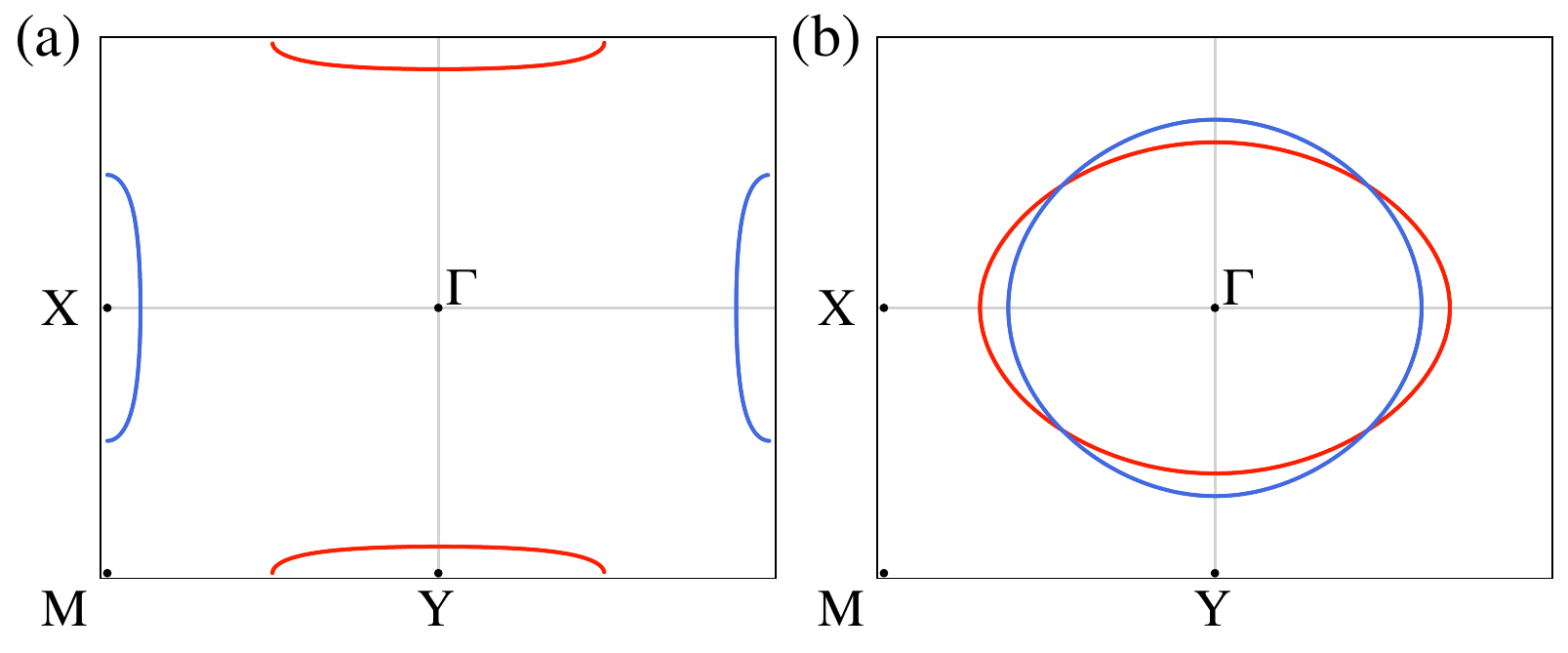}
    \caption{Fermi surfaces for spin-up (red) and spin-down (blue) for two different chemical potentials. (a) $\mu=0.5t_1$, and (b) $\mu=-0.5t_1$}
    \label{fig:FSs}
\end{figure}

One of the motivating questions for this work is whether self-energy effects will mask the bare spin splitting. 
To explore this, we consider the spectral functions for electron-phonon interactions and electron-magnon interactions separately in Fig.~\ref{fig:Spectral functions}. The corrections to the band structure due to electron-phonon interactions are relatively minor, as the spectral function is peaked and barely shifted from the original bands. The shifting of the bands due to electron-magnon interactions is, for the most part, minor, though most significant for the upper band around the $\Gamma$-point. The bare spin splitting remains well resolved in the spectral function, even though the self-energies do not have a negligible magnitude.

\begin{figure}[tb]
    \centering
    \includegraphics[width=\linewidth]{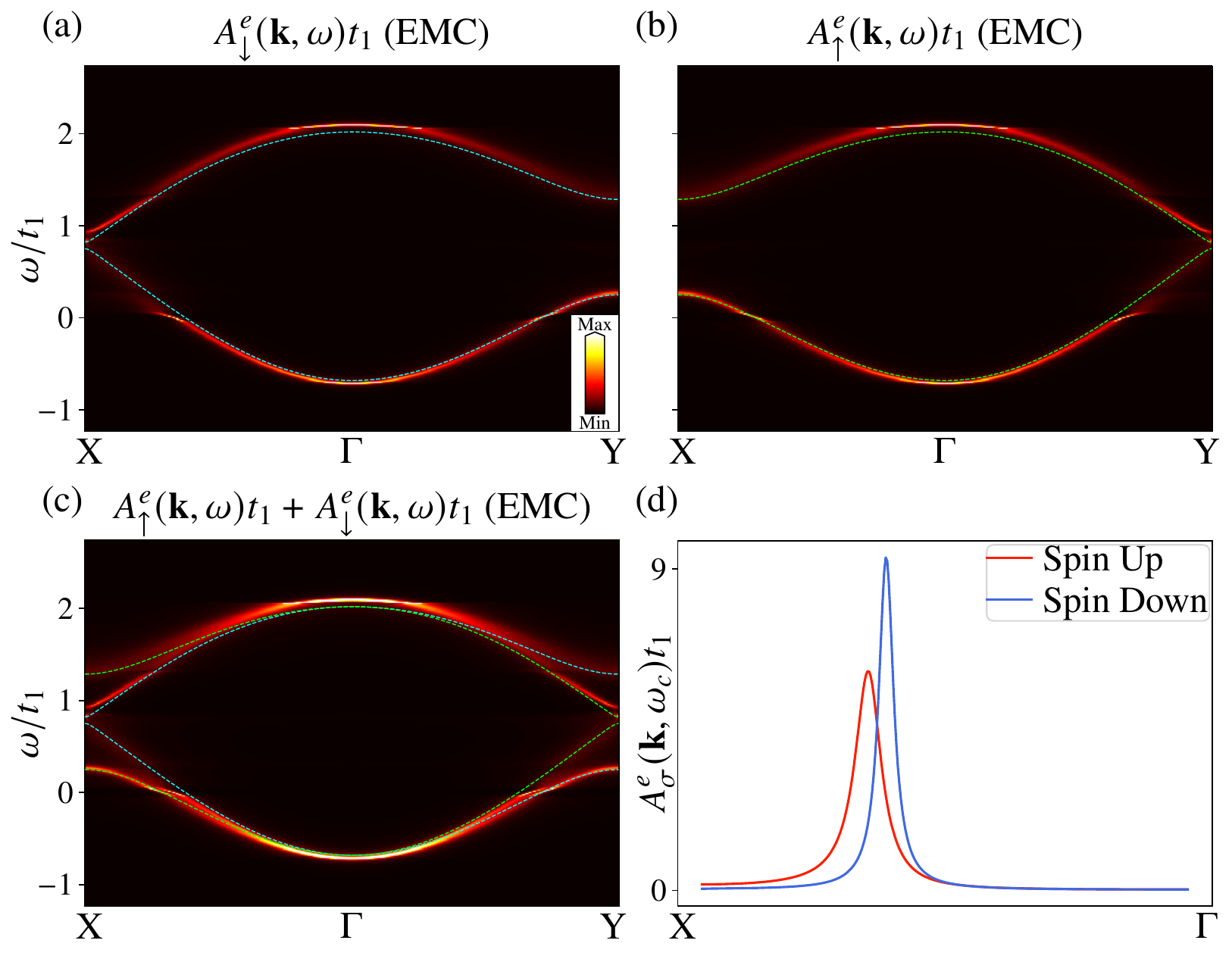}
    \caption{(a)-(c) The spectral function as a function of $\textbf{k}$ and $\omega$ along a path between the symmetry points in the BZ for the case of electron-magnon coupling. The chemical potential is set at $\mu=-0.5t$ and $T=20$~K. The color map is the same for all figures, with black corresponding to $A(\textbf{k},\omega)t=0$ and white $A(\textbf{k},\omega)t\geq30$. The original electron bands are shown as dashed blue (green) lines for spin-down (up). \textit{e} refers to \textit{elliptical}, as we consider the FS shown in \figref{fig:FSs}{b}. (a) Shows the case of spin-down electrons, (b) the case of spin-up, and (c) the sum of these. (d) Shows the spectral function for the two spins as a function of $\textbf{k}$ along the X-$\Gamma$ line for a fixed $\omega=\omega_c\equiv-6.7\times10^{-2}t$. Parameters used are given in Tab.~\ref{tab:Parameters}.}
    \label{fig:Spectral for ellipse}
\end{figure}

From Fig.~\ref{fig:SE} (where we have set $\mu = 0.5 t_1$ such that the FSs are arcs around the X and Y points, see \figref{fig:FSs}{a}), we see that interactions with magnons give different broadening for $\omega < 0$ and $\omega > 0$. Of particular interest is that the largest broadenings for the two spin species are at opposite sides of $\omega=0$. 
Thus, for electrons right below the Fermi level, we can distinguish the electron spins by only considering the width of the experimental data. 
It is important for such an analysis that a Fermi surface exists for both spin up and spin down in the same direction in momentum space. 
To achieve this, we consider the case in which both spins have an elliptical FS around the $\Gamma$-point, such that the bare bands of both spins cross the Fermi level in the same momentum direction, as shown in \figref{fig:FSs}{b}. Due to the $(C_{4z}|\mathcal{T})$ symmetry of the lattice, all effects flip in terms of spin upon rotating the momentum by $\pi/2$, which is why both spins must cross the Fermi level in the same direction to distinguish the spins based upon broadening.
We tune the chemical potential to lie in the lower band such that the FSs are closed contours (ellipses) around the $\Gamma$-point. To maintain spin-splitting and avoid van Hove singularities with such FSs, we set $\mu=-0.5 t_1$
(corresponding to a Fermi level at $-1t_1$ in \figref{fig:Bands}{a}). 
The spectral function for this case can be found in Fig.~\ref{fig:Spectral for ellipse}.
\figref{fig:Spectral for ellipse}{d} considers the case of $|\omega|$ being slightly larger than the magnon bandwidth, and we see, as expected, the spin-up band is broader than the spin-down band. Experimentally, we can use this to gain spin-resolved information without any spin filter. If two suspected spin-split bands are measured in ARPES on a candidate altermagnet, different broadening will be an indication of the opposite spin nature of the bands. At the same time, it is an indirect measurement of the magnon spin splitting.

We suggest that this spin-dependent lifetime of the electron quasiparticles can explain the experimental result from ARPES on the $d$-wave Lieb lattice altermagnet $\rm{Rb_{1-\delta} V_2 Te_2 O}$ \cite{FZhang}. Their Fig.~2 demonstrates different spectral broadening for the spin-split bands $\alpha$ and $\beta$ between the X- and M-points.

Considering Figs.~\ref{fig:Spectral functions} and \ref{fig:Spectral for ellipse}, we now discuss the more general lifetime effects in the full electron bandwidth. The most significant difference between magnon and phonon cases is that, for scattering off magnons, the electron bands become broadened for momentum values in the vicinity of the bare electronic band crossing, see Fig.~\ref{fig:Spectral functions} and \figref{fig:Bands}{a}. 
This is due to the van Hove singularity close to the crossings. 
The density of states diverges at flat bands, yielding a large number of electron states available to scatter in these regions. Hence, the spectral function will broaden \cite{Schrodi_ReplicaBand}. 
The difference in broadening in the phonon and magnon cases is due to the weaker electron-phonon coupling compared to the electron-magnon coupling. 
The same enhanced broadening can be seen even more clearly in Fig.~\ref{fig:Spectral for ellipse}, where there are shadow bands located
around the almost flat bands for opposite spin species. For example, in \figref{fig:Spectral for ellipse}{a}, the spin-down spectral function broadens along the X-$\Gamma$ line, which from \figref{fig:Bands}{a} corresponds to the van Hove singularity of spin-up electrons at the X-point. 
Despite the broadening due to these van Hove singularities, we see from \figref{fig:Spectral functions}{f} and \figref{fig:Spectral for ellipse}{c}, that the renormalized bands are still experimentally spin resolved.

When computing the results shown in Fig.~\ref{fig:Spectral functions} numerically, we are limited by computational power in the $\textbf{k}$-resolution. Because of this, we first calculated $\Sigma(\textbf{k},\omega)$ on a set of $40$ $\textbf{k}$-points between each symmetry point. To increase the resolution, we add $10$ additional $\textbf{k}$ values between the original points.
At these intermediate points, we assign the value from the nearest
$\Sigma(\textbf{k},\omega)$ from the original set of $\textbf{k}$ values, while recomputing $E_\textbf{k}$ at the refined values. The self-energy varies slowly as a function of momentum, such that this approximation gives indistinguishable results when considering plots as Fig.~\ref{fig:Spectral functions}.
The slow variation of the self-energy as a function of momentum when $|\omega|$ is greater than the boson bandwidth is because all bosons will contribute to scattering. In this case, $\mr{Im}\Sigma$ goes as the electronic density of states.
However, even when $|\omega|$ is smaller than the boson bandwidth, it is not possible to see any difference in the plots depicting the spectral function in terms of both momentum and energy. The approximation could have been visible in \figref{fig:Spectral for ellipse}{d}, but we do not apply this approximation 
in Fig.~\ref{fig:Spectral for ellipse} since the electron-magnon coupling is less computationally costly.

We have focused on a two-dimensional $d$-wave Lieb lattice altermagnet as a convenient model. We believe the main points of our results are quite general with respect to dimension, lattice structure, and angular momentum of the spin splitting. Hence, most of our results are also relevant for three-dimensional (3D) $g$-wave altermagnets, apart from the analysis of the magnetoelastic coupling, which is more complicated in 3D.

\section{Conclusion}
\label{sec:Conclusion}
We have presented a comprehensive analysis of many-body renormalization effects on the spin-split electronic structure of altermagnets, focusing on self-energy contributions arising from electron interactions with magnons, phonons, and magneto-elastic hybrid modes. 
By identifying a reasonable set of parameter regimes in which the splitting remains experimentally resolvable, we provide quantitative criteria for interpreting angle-resolved photoemission and tunneling experiments on altermagnetic materials.

A key outcome of our study is the pronounced asymmetry in magnon-induced broadening between the two spin channels near the Fermi surface, an effect absent in purely phononic contributions. The same conclusion holds when magnetoelastic coupling is included. This distinction underscores the unique role of spin-selective scattering in altermagnets and highlights the utility of linewidth analysis as a diagnostic for magnon-driven many-body processes. Moreover, we find that magneto-elastic coupling yields broadening behavior closely resembling that of the magnon-only case, indicating that the magnetic character of the hybrid modes dominates their impact on the electronic spectrum.

By incorporating the full temperature dependence of the self-energies, we further quantify how thermal fluctuations enhance quasiparticle decay and progressively reduce the visibility of spin splitting. These findings establish a microscopic framework for understanding lifetime effects in spin-split systems and offer guidance for the optimal conditions under which altermagnetic spin textures can be spectroscopically resolved. Overall, our work advances the theoretical foundation for interpreting experimental signatures of altermagnets and contributes to the broader effort to chart many-body phenomena in systems with unconventional spin symmetry.

\begin{acknowledgments}
We thank Anders C. Mathisen, Zhuoyuan Shi, Erlend Syljuåsen, and Xin-Liang Tan for helpful discussions.
KL and AS were supported by the Research Council of Norway (RCN) through its Centres of Excellence funding scheme, Project No.\ 262633, ``QuSpin'', RCN Project No.\ 323766, as well as COST Action CA21144  ``Superconducting Nanodevices and Quantum Materials for Coherent Manipulation". KM was supported by the DFG (SFB 1170) and the W{\"u}rzburg-Dresden Cluster of Excellence ct.qmat, EXC 2147 (Project-Id 390858490).
\end{acknowledgments}

\appendix
\section{Derivation of magnon-phonon coupling}
\label{app:Magnon-phonon coupling derivation}
Starting from the magnon-phonon interaction \eqref{eq:Magnon-phonon Hamiltonian}, we derive the quantized interaction shown in \eqref{eq:Diagonal magnon-phonon Hamiltonian}.
The first step is to split the Hamiltonian of \eqref{eq:Magnon-phonon Hamiltonian} as $H_\mathrm{mp}=H_\mathrm{mp}^A+H_\mathrm{mp}^B$, where $H_\mathrm{mp}^{A(B)}$ contains the terms where $i\in L_A(L_B)$. The sets $L_A$ and $L_B$ denote the sublattices for spin-up and spin-down lattice sites, respectively. With this, we have
\begin{align}
\label{eq:H-A def}
    &H_\mathrm{mp}^A=\sum_{i\in L_A}\sum_{j=1}^{J_A}\dfrac{\kappa}{2|\boldsymbol{\delta}^A_j|}\dfrac{S_{iA}^{\raisebox{0.5ex}{$\scriptstyle z$}}}{S}\left(\textbf{S}_{iA}\cdot\Hat{\boldsymbol{\delta}}_{j}^A\right)(u_{iA}^z-u_{jB}^z),\\
    &H_\mathrm{mp}^B=\sum_{i\in L_B}\sum_{j=1}^{J_B}\dfrac{\kappa}{2|\boldsymbol{\delta}^B_j|}\dfrac{S_{iB}^{\raisebox{0.5ex}{$\scriptstyle z$}}}{S}\left(\textbf{S}_{iB}\cdot\Hat{\boldsymbol{\delta}}_{j}^B\right)(u_{iB}^z-u_{jA}^z),
\end{align}
where $J_{A(B)}$ is the number of nearest neighbors belonging to sublattice $B(A)$. $\boldsymbol{\delta}^{A(B)}_j$ is the vector from a site on sublattice $A(B)$ to its nearest neighbor on the other magnetic sublattice, and the hat indicates that the vector is a unit vector.
We focus on $H_\mathrm{mp}^A$, as the derivation for $H_\mathrm{mp}^B$ is similar. From the lattice structure in Fig.~\ref{fig:Lieb lattice}, we see that for a spin-up site, the nearest spin-down neighbors
are located at $\boldsymbol{\delta}_{j}^A\in a\left\{(1,-1),(1,1), (-1,1), (-1,-1)\right\}$. Such that $|\boldsymbol{\delta}_{j}^A|=\sqrt{2}a\,\forall\,j$.
In \eqref{eq:H-A def}, the term proportional to $u_{iA}^z$ depends on \textit{j} only through $\Hat{\boldsymbol{\delta}}_{j}^A$, which, due to the inversion symmetric lattice structure, will cancel when summing over \textit{j}. This simplifies $H_\mathrm{mp}^A$ to
\begin{equation}
    H_\mathrm{mp}^A=-\dfrac{\kappa}{2\sqrt{2}aS}\sum_{i\in L_A}\sum_{j=1}^{J_A}S_{iA}^{{\raisebox{0.5ex}{$\scriptstyle z$}}}\left(\textbf{S}_{iA}\cdot\Hat{\boldsymbol{\delta}}_{j}^A\right)u_{jB}^z.
\end{equation}
We proceed by expanding the dot product and rewriting the spin operators as ladder operators by a HP transformation,
\begin{equation}
\label{eq:HP A}
S_i^+\approx \sqrt{2S}\,a_i,     \colgap
S_i^-\approx \sqrt{2S}\,a_i^{\dagger}, \colgap
S_{i}^{{\raisebox{0.5ex}{$\scriptstyle z$}}}\approx S-a_i^\dagger a_i, \colgap i\in L_A,\\
\end{equation}
where $S^{\pm}\equiv S^x\pm iS^y$, and $a_i$ is a magnon annihilation operator.
Inserting this into the Hamiltonian yields
\begin{equation}
    H_\mathrm{mp}^A=-\dfrac{\kappa\sqrt{S}}{4a}\sum_{i\in L_A}\sum_{j=1}^{J_A}\left[a_i\left(\Hat{\delta}_{jx}^A-i\Hat{\delta}_{jy}^A\right)u_{jB}^z+\mathrm{H.c.}\right],
\end{equation}
where we only consider terms linear in magnon operators. The next step is to quantize the displacements by utilizing \eqref{eq:Displacement phonon}, 
\begin{equation}
\begin{aligned}
    H_\mathrm{mp}^A=\dfrac{\kappa\sqrt{S}}{4a}\sum_{i\in L_A}\sum_{\textbf{q}\lambda}\sqrt{\dfrac{1}{2M_B\omega_{\textbf{q}\lambda}}}\left(\boldsymbol{e}_\textbf{q}^{(\lambda)}\right)^z_B&\\
    \times\left[a_i^{(\mr{m})}\left(a_{\textbf{q}\lambda}^{(\mr{p})\dag}+a^{(\mr{p})}_{-\textbf{q}\lambda}\right)e^{-i\textbf{R}_i\cdot\textbf{q}}\Gamma^A(\textbf{q})+\mathrm{H.c.}\right]&,
\end{aligned}
\end{equation}
where we distinguish between magnon and phonon operators with the labels (m) and (p). The factor $\Gamma^A(\textbf{q})$ is defined as
\begin{equation}
    \begin{aligned}
        &\Gamma^A(\textbf{q})\equiv-\sum_{j=1}^{J_A}\left(\Hat{\delta}_{jx}^A-i\Hat{\delta}_{jy}^A\right)e^{i\boldsymbol{\delta}_j^A\cdot\textbf{q}}\\
        &=-2\sqrt{2}\left[\cos(aq_x)\sin(aq_y)+i\sin(aq_x)\cos(aq_y)\right].
    \end{aligned}
\end{equation}
We now insert the FT $a_i^{(\mr{m})}=\frac{1}{\sqrt{N}}\sum_\textbf{q}a^{(\mr{m})}_\textbf{q}e^{-i\textbf{R}_i\cdot\textbf{q}}$ and perform the sum over \textit{i}, which gives
\begin{equation}
\begin{aligned}
    H_\mathrm{mp}^A=\dfrac{\kappa\sqrt{S}}{4a}\sum_{\textbf{q}\lambda}\sqrt{\dfrac{1}{2M_B\omega_{\textbf{q}\lambda}}}\left(\boldsymbol{e}_\textbf{q}^{(\lambda)}\right)^z_B&\\
    \times\left[a^{(\mr{m})}_\textbf{q}\left(a_{\textbf{q}\lambda}^{(\mr{p})\dag}+a^{(\mr{p})}_{-\textbf{q}\lambda}\right)\Gamma^A(\textbf{q})+\mathrm{H.c.}\right]&,
\end{aligned}
\end{equation}
The last step to derive \eqref{eq:Diagonal magnon-phonon Hamiltonian} is to substitute the sublattice HP operators with the diagonal Bogoliubov operators given by \eqref{eq:Bogoliubov}. The procedure for $H_\mathrm{mp}^B$ is similar, using the HP transformations in \eqref{eq:HP A}, and \eqref{eq:HP B}.
\begin{alignat}{3}
\label{eq:HP B}
S_j^+&\approx \sqrt{2S}\,b_j^\dagger, &\colgap
S_j^-&\approx \sqrt{2S}\,b_j,       &\colgap
S_{j}^{{\raisebox{0.5ex}{$\scriptstyle z$}}}&\approx b_j^\dagger b_j-S, \colgap  j\in L_B,
\end{alignat}

\section{Diagonalization of a quadratic bosonic Hamiltonian}
\label{app:Diagonalization of quadratic bosonic Hamiltonian}
In \eqref{eq:Simple H}, we considered a structure of the Nambu arrays to simplify the Hamiltonian, namely \eqref{eq:Nambudagger} and \eqref{eq:Nambu}. The following procedure for diagonalizing a quadratic bosonic Hamiltonian requires a different structure of the Nambu array, where all annihilation operators appear in the first elements, while all respective creation operators appear in the last elements. To achieve the same results with the simplified Hamiltonian, we have to perform a permutation matrix $P$, which transforms the Hamiltonian as $P^{-1}HP$ so that $\Hat{\boldsymbol{\Phi}}_\textbf{q}$ acquires the standard form with all annihilation operators first, followed by the respective creation operators. For example $P_{36}=1$ as we move 
$\alpha^\dag_{-\textbf{q}}$ from the third position in $\Hat{\boldsymbol{\Phi}}_\textbf{q}$ to the fifth position in $\boldsymbol{\Phi}_\textbf{q}$.
Consider the general quadratic bosonic Hamiltonians in the form
\begin{align}
    \mathcal{H}=&\dfrac{1}{2}\boldsymbol{\Phi}^\dag H\boldsymbol{\Phi},\\
    \label{eq:Bosonic vector dagger}
    \boldsymbol{\Phi}^\dag=&\begin{pmatrix}
        b_1^\dag&\ldots&b_N^\dag&b_1&\ldots&b_N
    \end{pmatrix},\\
    \label{eq:Bosonic vector}
    \boldsymbol{\Phi}=&\begin{pmatrix}
        b_1&\ldots&b_N&b_1^\dag&\ldots&b_N^\dag
    \end{pmatrix}^T,\\
    H=&\begin{pmatrix}
        \Xi&\Delta\\
        \Delta^*&\Xi^*
    \end{pmatrix},
\end{align}
where we have $N$ different bosonic operators $b_i$. The bosonic commutation relations in terms of the Nambu arrays $\boldsymbol{\Phi}$ are given by
\begin{align}
    \left[\Phi_n,\Phi_m^\dag\right]&=g_{nm},\\
    \left[\Phi_n,\Phi_m\right]&=-\left[\Phi_n^\dag,\Phi_m^\dag\right]=i\left(\Sigma_y\right)_{nm},
\end{align}
where $g\equiv\sigma_z\otimes1_N$ with $\sigma_z$ being the third Pauli matrix and $1_N$ is the $N\times N$ identity matrix, likewise $\Sigma_y\equiv\sigma_y\otimes1_N$. We now seek a way to diagonalize such Hamiltonians. This is done by applying a Bogoliubov-Valatin transformation \cite{Colpa78}, which introduces new bosonic quasiparticles,
\begin{equation}
    \label{eq:Magnon-phonon quasiparticles}
    \boldsymbol{\psi}^\dag=\begin{pmatrix}
        \psi^\dag_1&\ldots&\psi^\dag_N&\psi_1&\ldots&\psi_N
    \end{pmatrix},
\end{equation}
which also satisfy the condition $\left[\psi_n,\psi_m^\dag\right]=g_{nm}$. The transformation matrix, \textit{T}, connects the new and old operators by $\boldsymbol{\Phi}=T\boldsymbol{\psi}$. Substituting this into the Hamiltonian gives
\begin{equation}
    \label{eq:Diagonalization Bosonic Hamiltonian}
    \mathcal{H}=\dfrac{1}{2}\boldsymbol{\Phi}^\dag H\boldsymbol{\Phi}=\dfrac{1}{2}\boldsymbol{\psi}^\dag T^\dag HT\boldsymbol{\psi}=\dfrac{1}{2}\boldsymbol{\psi}^\dag E\boldsymbol{\psi},
\end{equation}
where $E$ is a diagonal matrix containing the energies. By the constraint that the new quasiparticles must satisfy the bosonic commutation relations, we find the following constraint on the transformation matrix,
\begin{equation}
    g_{nm}=\left[\Phi_n,\Phi_m^\dag\right]=\sum_{ij}T_{ni}\left[\psi_i,\psi_j^\dag\right]T^\dag_{jm}=\left(TgT^\dag\right)_{nm},
\end{equation}
meaning $T^\dag gT=g\Leftrightarrow T^\dag=gT^{-1}g$. Matrices satisfying this constraint are referred to as para-unitary \cite{Colpa78}. Inserting this relation into the condition $E=T^\dag HT$ from \eqref{eq:Diagonalization Bosonic Hamiltonian}, yields
\begin{equation}
    T^{-1}\left(gH\right)T=gE.
\end{equation}
Hence, we obtain the transformation matrix by diagonalizing the \textit{non-hermitian} matrix $gH$ by following the algorithm proposed by Colpa in Ref.~\cite{Colpa78} 
See App.~B of Ref.~\cite{SkyrmionMS} for a description of the method including the momentum dependence.

\section{Numerical algorithm to solve for nodal lines}
\label{App:Nodal lines}
We utilize Python for the numerical computations and the following algorithm to obtain the nodal lines required to calculate \eqref{eq:Self-energy imaginary part}.

\begin{enumerate}
    \item Create a grid over a \emph{padded} 1BZ to better capture curves close to the 1BZ edges. We choose the padding such that the grid extends from $-1.02$X to $1.02$X along \textit{x}-direction and likewise along \textit{y}. We use $500$ points along each axis, resulting in a total of $25\times10^4$ points. Then shift every other row by $d/2$, where \textit{d} is the distance between neighboring points

    \item Perform a Delaunay triangulation to form triangles with these points as vertices. To avoid triangles with long edges near the boundary of the padded 1BZ, filter out triangles with edges longer than $\sqrt{10}d$.

    \item Evaluate the electron and phonon energies at each point in the grid.

    \item Loop over all combinations of electron band, phonon mode, and $\chi$ to calculate $\xi_\textbf{q}$. Check if the lowest and largest value of $\xi_\textbf{q}$ have opposite signs. If they do, check that a sufficient number of points satisfy $|\xi_\textbf{q}|<tol$, where we have set $tol=0.3$~eV and require at least $500$ such points. This is to ensure we actually get a curve we can integrate over, not just a few points. 

    \item To robustly find the nodal curves, the contouring step will need some triangles with a larger value of $|\xi_\textbf{q}|$. Only including small $|\xi_\textbf{q}|$ will make it difficult for the algorithm to grasp what is a good enough accuracy. So we choose to include all triangles with a vertex satisfying $|\xi_\textbf{q}|<5tol$.

    \item The algorithm to solve for the nodal curves takes in the triangulation and the value of $\xi_\textbf{q}$ at each vertex. Then, if a vertex in a given triangle has an opposite sign to the other two vertices, it uses linear interpolation along the edges of the triangle to find the nodal point. Then draw a line between the two nodal points in a given triangle and connect this with the lines from all the other triangles. Specifically, we use the class \textit{TriContourGenerator} from \textit{Matplotlib} \cite{Matplotlib} in this step. 
    
    \item Because we started from a padded 1BZ, we then mask out all points lying outside the original 1BZ boundary. This might split some curves. To regroup them into continuous curves, we apply the \textit{DBSCAN} clustering algorithm.
    
    \item To ensure that the points along each curve are ordered continuously (as required for line integration), we locate the two points in a given curve that have the largest mutual distance. If these points are not already the first and last entries in the array, we \textit{roll} the indices so that this pair becomes the endpoints. 

    \item The last step is to ensure that the distance between each point in a given curve is equidistant. To do this, compute the cumulative arc length and use that as a parameterization of the curve. Sample evenly spaced arc-length positions and linearly interpolate the $(x,y)$ coordinates to form a new evenly spaced curve. Repeat this process until the ratio between the longest and smallest distance is less than a tolerance \textit{ratio}, where we used \textit{ratio} $=1.001$.
\end{enumerate}

\section{Determining magnon species}
\label{app:Determining magnon species}
In the main text, we have seen the consequence when different magnons couple to electrons with different spins. We will here give an analysis of why that happens. In Sec.~\ref{sec:Self-energy}, the condition for integral paths is given by
\begin{equation}
    \omega-E_{\textbf{k}+\textbf{q}}+\chi\omega_\textbf{q}=0,
\end{equation}
where we have simplified the notation by dropping some indices. From this, the electron energy is conditioned by
\begin{equation}
    E_{\textbf{k}+\textbf{q}}=s|\omega|+\chi|\omega_\textbf{q}|,
\end{equation}
where \textit{s} gives the sign of $\omega$ and the magnon energy is always positive. We now insert this expression into the hyperbolic factor in \eqref{eq:Self-energy imaginary part} and use the general relation
\begin{equation}
    \tanh(x+y)=\frac{\tanh(x)+\tanh(y)}{1+\tanh(x)\tanh(y)}
\end{equation}
to rewrite the hyperbolic factor as
\begin{equation}
\begin{aligned}
    &\coth(\frac{\omega_\textbf{q}}{2k_BT})-\chi\tanh(\frac{E_{\textbf{k}+\textbf{q}}}{2k_BT})\\
    =&\coth(\frac{|\omega_\textbf{q}|}{2k_BT})-\chi\tanh(\frac{s|\omega|+\chi|\omega_\textbf{q}|}{2k_BT})\\
    =&\coth(\frac{|\omega_\textbf{q}|}{2k_BT})-\dfrac{s\chi\tanh(\frac{|\omega|}{2k_BT})+\tanh(\frac{|\omega_\textbf{q}|}{2k_BT})}{1+s\chi\tanh(\frac{|\omega|}{2k_BT})\tanh(\frac{|\omega_\textbf{q}|}{2k_BT})}.
\end{aligned}
\end{equation}
Analyzing this expression in terms of $s$ and $\chi$, we see that the largest contribution from the numerator comes when $s\chi=-1$. Since the arguments of $\tanh$ in the denominator are all positive, $\tanh$ will be between $0$ and $1$, such that the denominator also gives the largest contribution when $s\chi=-1$. Hence, the sign of $\omega$ determines the dominating sign of $\chi$, and thus which magnon mode contributes to the self-energy.

\begin{table}[htbp]
    \centering
    \caption{List of parameters}
    \begin{tabular}{l c @{\hspace{1cm}} l c}
        \hline
        \hline
        \textbf{Parameter} & \textbf{Value} &
        \textbf{Parameter} & \textbf{Value} \\
        \hline
        $t_1$&$600$~meV & $M_3$&$55$~u\\
        $t_2$&$300$~meV & $\eta_\mr{IP}$&$-4.0$~N/m\\
        $t_3$& $75$~meV & $\rho_\mr{IP}$&$-5.7$~N/m\\
        $s_x$&$0.6$~\AA & $\gamma_\mr{IP}$&$-8.0$~N/m\\
        $s_y$&$0.6$~\AA & $\eta_\mr{OOP}$&$-2.0$~N/m\\
        $J_{\mathrm{sd}}$&$150$~meV & $\rho_\mr{OOP}$&$-2.8$~N/m\\
        $J_\mr{AB}$&$5.8$~meV & $\gamma_\mr{OOP}$&$-4.0$~N/m\\
        $J_\mr{nm}$& $-2.3$~meV & $\varepsilon_\mr{nm}$&$-4000$~meV\\
        $J_\mr{d}$&$-1.2$~meV &  $a$&$1.0$~\AA\\
        $\mu$& $300$~meV & $K_z$&$0.6$~meV \\
        $M_1$&$55$~u & $\kappa$&$15$~meV\\
        $M_2$&$16$~u & $S$&$1$\\
        \hline
        \hline
    \end{tabular}
    \label{tab:Parameters}
\end{table}

\section{Numerical values}
\label{app:Numerical values}
Some of lattice parameters are based on density functional theory (DFT) calculations of the altermagnetic material $\mr{La_2 O_3 Mn_2 Se_2}$ done by Ref.~\cite{GRSMV25}. As our model is two-dimensional, we use the values for a layer of $\mr{Mn^{2+}}$, $\mr{O^{2-}}$, and $\mr{
Se^{2-}}$. To map this structure to our Lieb lattice, we neglect the $\mr{Se^{2-}}$ atoms. The parameters we use based upon $\mr{La_2 O_3 Mn_2 Se_2}$ are the ion masses, the on-site energy of the non-magnetic site $(\varepsilon_\mr{nm})$, and the hoppings $t_2$ and $t_3$.

With a value of the lattice constant $a$ corresponding to the material $\mr{La_2 O_3 Mn_2 Se_2}$ \cite{GRSMV25}, the self-energy effects due to phonons will be negligible compared to those due to the magnons. Thus, we set the lattice constant to $a=1$~\AA, as the electron-phonon coupling, \eqref{eq:Electron-phonon coupling}, goes as $a^{-1}$. That way, we emulate a case where the electron-phonon coupling is strong enough to give comparable self-energy effects to magnons. That could be the case in another material.

The parameters associated with the magnons and phonons are chosen to give reasonable band structures and bandwidths of about $1/50$ of the electronic bandwidth. The standard deviations $s_x$ and $s_y$ are set equal and such that $t_3/t_2=1/4$ as proposed by Ref.~\cite{GRSMV25}. The value of $t_1$ is then set by the standard deviations, the value of $t_2$ and the value of $t_3$. We set the value of the exchange interaction $J_\mr{sd}$ to achieve a spin splitting within the expected range of altermagnets \cite{Smejkal-Sinova-Jungwirth, Smejkal2022Dec}.

Table~\ref{tab:Parameters} contains all parameter values used in the paper if not otherwise specified.

\bibliography{Refs.bib}
\end{document}